\documentclass[11pt]{article}
\pdfoutput=1
\usepackage{jheppub}
\usepackage[T1]{fontenc} 

\usepackage{amssymb}
\usepackage{amsfonts}
\usepackage{amsbsy}
\usepackage{amsmath}
\usepackage{amsthm}
\usepackage{graphicx}
\usepackage{epstopdf}
\usepackage{multirow}
\usepackage{latexsym}
\usepackage{array}

\input cyracc.def 
\newfont{\twelvecyr}{wncyr10 at 12pt}

\def\Z{\mathbb{Z}}
\def\F{\mathbb{F}}
\def\Q{\mathbb{Q}}
\def\C{\mathbb{C}}

\def\P{\mathbb{P}}

\def\n3a{t}

\def\ge{{\mathfrak{e}}}
\def\gso{{\mathfrak{so}}}
\def\gsu{{\mathfrak{su}}}
\def\gsp{{\mathfrak{sp}}}
\def\gf{{\mathfrak{f}}}
\def\gg{{\mathfrak{g}}}

\title{Non-Higgsable clusters for 4D F-theory models}

\author[1]{David R.  Morrison}
\author[]{and}
\author[2]{Washington Taylor}

\affiliation[1]{Departments of Mathematics and  
Physics\\ University of California, Santa Barbara\\ Santa Barbara, CA 93106, USA}
\affiliation[2]{Center for Theoretical Physics\\
Department of Physics\\
Massachusetts Institute of Technology\\
77 Massachusetts Avenue\\
Cambridge, MA 02139, USA}

\emailAdd{{\tt drm} {\rm at} {\tt math.ucsb.edu}}
\emailAdd{{\tt wati} {\rm at} {\tt mit.edu}}

\preprint{UCSB Math  2014-38, MIT-CTP-4582}

\abstract{
We analyze non-Higgsable clusters of gauge groups and matter that can
arise at the level of geometry in 4D F-theory models.  Non-Higgsable
clusters seem to be generic features of F-theory compactifications,
and give rise naturally to structures that include the
nonabelian part of the standard model gauge group and certain specific
types of potential dark matter candidates.  In particular, there are
nine distinct single nonabelian gauge group factors, and only five
distinct products of two nonabelian gauge group factors with matter,
including $SU(3) \times SU(2)$, that can be realized through 4D
non-Higgsable clusters.  There are also more complicated
configurations involving more than two gauge factors; in particular,
the collection of gauge group factors with jointly charged matter can
exhibit branchings, loops, and long linear chains.  }

\begin{document}

\maketitle

\flushbottom

\section{Introduction}

Many supersymmetric string
theory compactifications contain ``non-Higgsable'' gauge
groups that cannot be broken by charged matter in a way that preserves
supersymmetry.  In the simplest cases, the non-Higgsable gauge group
is a single simple factor such as $SU(3), SO(8),$ or $E_8$ under which
there are no charged matter fields.  Such string vacua have long been
known to arise in heterotic string compactifications, and in many
cases have dual F-theory descriptions \cite{Vafa-F-theory,
  Morrison-Vafa-I, Morrison-Vafa-II}; a simple set of examples are
given by 6D supergravity theories arising from  heterotic
compactifications on K3 and dual F-theory compactifications on Hirzebruch
surfaces $\F_m$.  For example, a non-Higgsable $E_8$ arises in the
$E_8 \times E_8$
heterotic theory when all of the 24 instantons needed for tadpole
cancellation in the 10D theory are placed in one of the two $E_8$
heterotic factors, corresponding on the F-theory side to
compactification on $\F_{12}$.  While in the simplest cases there is
no charged matter, there are also cases where a gauge group is
non-Higgsable even in the presence of charged matter.  For example, in
a 6D heterotic compactification where the numbers of instantons in the
two $E_8$ factors are 5 and 19, corresponding on the F-theory side to
a compactification on $\F_7$, there is a non-Higgsable gauge group
$E_7$ carrying a half hypermultiplet in the ${\bf 56}$ representation.
This matter cannot be Higgsed in the low-energy theory since the
D-term constraints cannot be satisfied by matter in a single real
representation.

In heterotic constructions that use smooth bundles over smooth
Calabi-Yau manifolds the  non-Higgsable gauge groups contain only
a single simple factor.  In F-theory, however, many geometries give
rise to non-Higgsable gauge groups with multiple factors and jointly
charged matter.  In \cite{clusters}, we performed a systematic
analysis of the possible non-Higgsable structures that can arise in 6D
F-theory compactifications, and identified all possible
``non-Higgsable clusters'' of gauge group factors connected by jointly
charged matter that cannot be broken by Higgsing.  The analysis was
carried out by looking at configurations of intersecting curves on the
(two complex dimensional) base surface, with each curve having a
negative self-intersection.  The complete list of non-Higgsable gauge
groups in 6D F-theory models contains, in addition to the single group
factors $SU(3),SO(8), F_4, E_6, E_7,$ and $E_8$, the two product
groups $G_2 \times SU(2)$ and $SU(2) \times SO(7) \times SU(2)$, with
non-Higgsable matter jointly charged under the adjacent factors in
each gauge group\footnote{The non-Higgsable structure imposed from
  geometry determines only the gauge algebra, so that these groups may
  in principle be reduced through a quotient by a finite subgroup in
  some cases.}.

In this paper we initiate a systematic analysis of non-Higgsable
clusters for 4D F-theory models.  
Non-Higgsable clusters give rise to gauge groups and matter at generic
points in the moduli spaces of Calabi-Yau  fourfolds
over many bases that can be
used for F-theory compactification.  From our current understanding of
the space of elliptically fibered Calabi-Yau manifolds, it seems that
in fact the vast majority of F-theory compactifications will have such
structure.  Non-Higgsable clusters can give rise to the nonabelian
part of the standard model \cite{ghst}, as well as decoupled
or weakly interacting sectors
that have a natural possible interpretation as dark matter.  In
\S\ref{sec:applications}, we comment on some aspects of these
constructions that may be relevant to phenomenology.

There are several issues that make the analysis of non-Higgsable
clusters, and F-theory vacua in general, more complex for
four-dimensional models than for six-dimensional models.  In six
dimensions the geometric complex structure moduli space of an
elliptically fibered Calabi-Yau threefold matches with a continuous
moduli space of flat directions in the corresponding 6D supergravity
theory, so that there is a close correspondence between the structure
of the physical theory and the geometric data of F-theory (see for
example \cite{KMT-II}).  In four dimensions this connection is
obscured by the presence of a superpotential that lifts some of the
flat directions.  Viewing F-theory as dual to a limit of M-theory, the
superpotential is produced by G-flux on a Calabi-Yau fourfold (see
\cite{Denef-F-theory} for an introductory review).  There are also
additional degrees of freedom on the world-volume of IIB seven-brane
configurations that are as yet not well understood or incorporated into the
F-theory context.  Even for perturbative ({\it e.g.} $SU(N))$
seven-brane stacks, off-diagonal excitations of the world-volume
adjoint scalar fields encode expansion of D$p$-branes into
higher-dimensional D$(p + 2k)$-branes \cite{WT-Trieste, tv-0, tv-p,
  Myers}; these degrees of freedom are not encompassed in the complex
structure moduli of the elliptically fibered Calabi-Yau used for
F-theory and have been studied in that context as ``T-branes''
\cite{Donagi-Katz-Sharpe, T-branes, Donagi-Wijnholt-gluing, Anderson-Heckman-Katz}.  While
such excitations can be 
described locally, unlike for the complex structure degrees of freedom
in F-theory which have a global characterization in terms of
Weierstrass models there is no analogous general global formulation
of the full set of open string degrees of freedom associated with
perturbative brane configurations on a general compact space (see {\it
  e.g.} \cite{Douglas-curved, D-brane-actions} for some initial
efforts in this direction, and 
\cite{Ferrari, Collinucci-Savelli-1, Collinucci-Savelli-2} for more recent
developments and further references).  For nonperturbative seven-brane
configurations associated with exceptional groups, the open string
dynamics is even less transparent from the F-theory complex structure
point of view.  

The effects of G-flux and additional degrees of freedom can not only
lift flat directions in the moduli space, but can also modify the
spectrum of the theory.  On the one hand, the potential produced by
G-flux can drive the theory to a point of enhanced symmetry, while on
the other hand flux in the world-volume fields on a set of
seven-branes can also break the apparent geometric symmetry to a
smaller group.  G-flux also affects the matter spectrum of the theory,
and can give rise to chiral matter although the underlying Calabi-Yau
geometry in the F-theory picture describes only non-chiral (4D ${\cal
  N} = 2$) matter.  Some of these issues are discussed in more detail
in \cite{ghst}.  Though there has been substantial work on various
aspects of G-flux in 4D F-theory compactifications (see for example
\cite{G-flux01,G-flux02,G-flux03,G-flux04,G-flux05,G-flux06,G-flux07}), there is still no completely general way of analyzing
these effects in an arbitrary 4D compactification.  In this paper, we
focus only on the underlying geometry of the F-theory
compactification, in particular on the continuous moduli space of
complex structures for a given elliptically fibered Calabi-Yau
fourfold parameterized by a Weierstrass model.  When we refer to the
\emph{geometric gauge group} and \emph{geometric matter}, we refer
only to the gauge group and non-chiral matter associated with the
singularities of the Weierstrass model.  This analysis thus gives only
a first-order picture of the space of possibilities that can exist in
complete F-theory models.  To determine the actual physical gauge
group and matter the further incorporation of G-flux effects is
necessary, and we leave this further analysis to future work.  Another
complication in the analysis of 4D F-theory models is the presence of
codimension three loci where the Weierstrass coefficients $f, g$
vanish to degrees $(4, 6)$.  As we discuss in the next section, it is
not yet understood whether such singularities pose a problem for
consistency of 4D F-theory models, and we include vacua with such loci
in the analysis here.

After a brief review of some basic
aspects of F-theory in Section \ref{sec:review}, we begin in Section
\ref{sec:local} with a general set of
formulae that can be used to give a
lower bound for the orders of vanishing of the Weierstrass coefficients $f$ and
$g$ over any given divisor in a
complex threefold base.  These formulae control the local singularity
structure of the Weierstrass model and determine the factors that can
appear in a non-Higgsable cluster.  In Section \ref{sec:6D}, as a
warm-up exercise we use a simplified version of the local divisor
formulae to describe non-Higgsable clusters in 6D theories, and
reproduce the results of \cite{clusters} in a simple and direct way.
We then proceed
in \S\ref{sec:4D-single}
and the following sections to analyze the local structure of 4D clusters using
the general formulae.  We find a rich range of behavior, including
branchings, loops, and long linear chains of connected gauge group
factors. 
We conclude in \S\ref{sec:conclusions} 
with a discussion of some of the possible applications of 4D
non-Higgsable clusters.

\section{Review of F-theory basics} 
\label{sec:review} 

Here we summarize a few of the basic features of F-theory that are
central to the analysis of this paper.  More comprehensive reviews can
be found in \cite{Morrison-TASI, Denef-F-theory, WT-TASI}.

We consider F-theory as a nonperturbative formulation of type IIB
string theory.  A supersymmetric F-theory compactification to $10-2n$
dimensions is defined by a complex $n$-fold base $B_n$ that supports
an elliptic fibration with section $\pi: X \rightarrow B_n$ where the
total space $X$ is a Calabi-Yau $(n+1)$-fold.  The data of such an elliptic
fibration can be described by  a Weierstrass model
\cite{Nakayama}
\begin{equation}
 y^2 = x^3+ f x+ g \,, 
\label{eq:Weierstrass}
\end{equation}
where $f, g$ are sections of line bundles ${\cal O}(-4K), {\cal O}
(-6K)$ over the base $B_n$, with $- K$ the anti-canonical class on
$B_n$.  The Weierstrass parameters $f$ and $g$ can be described in
terms of polynomials of fixed degrees in a local coordinate system on
$B_n$.  The (geometric) gauge group of the corresponding supergravity
theory is determined by the codimension one singularity structure of
the Weierstrass model, where the discriminant $\Delta = 4f^{3}+27g^2$
vanishes.  When $f, g,$ and $\Delta$ vanish to certain  orders on a
divisor (\emph{i.e.}, a codimension one algebraic subspace) then the
total space of the elliptic fibration is singular, and can be viewed
as a degenerate limit of a smooth Calabi-Yau manifold; in IIB language the singularities can be interpreted in
terms of coincident seven-branes (but of more general types than
occur in the perturbative IIB string).  In either picture, the physical
result is the appearance of a nonabelian gauge symmetry in the
supergravity theory.  The classification of codimension one
singularities, following Kodaira \cite{Kodaira}, is listed in Table~\ref{t:Kodaira},
along with the resulting gauge algebra factors (which are inferred
from gauge symmetry enhancement in M-theory \cite{WitMF,Aspinwall:2000kf}).  
In some cases, for
compactifications to six dimensions or fewer the gauge group depends
not only on the orders of vanishing of $f, g, \Delta$, but also on
the more detailed monodromy structure of the singularity locus
\cite{Kodaira, Morrison-Vafa-II, Bershadsky-all, Morrison-sn, Grassi-Morrison-2}.  The
(geometric) matter content of the theory is determined by the
codimension two singularity locus on the base.  In simple cases, the
representation content of the matter is  determined in a simple
fashion
from the
enhancement of the Kodaira singularity type on the singular
codimension two locus \cite{Bershadsky-all, Katz-Vafa}, but more complicated matter representations can
also arise.  A complete dictionary between codimension two
singularities and matter representations has not yet been developed,
though a number of recent works have made progress in this direction
\cite{Grassi-Morrison, mt-singularities, Esole-Yau, Grassi-Morrison-2,
  Esole-Yau-II, Lawrie-Sakura, Grassi-hs-1, Hayashi-lms, Grassi-hs-2,
  Esole-sy}.  For the purposes of this paper, the most relevant
fact is that, in general, matter arises at codimension two loci within
codimension one divisors carrying gauge group factors, where the
degrees of vanishing of $f, g$, and/or $\Delta$ are enhanced.  In particular, when two
divisors each carry a gauge group factor, and they intersect along a
codimension two locus (a set of points in the case of 6D
compactifications, or a complex curve in the case of 4D
compactifications), then there is generally (geometric) matter that
carries a charge under both of the gauge group factors.

\begin{table}
\begin{center}
\begin{tabular}{|c |c |c |c |c |c |}
\hline
Type &
ord ($f$) &
ord ($g$) &
ord ($\Delta$) &
singularity & nonabelian symmetry algebra\\ \hline \hline
$I_0$&$\geq $ 0 & $\geq $ 0 & 0 & none & none \\
$I_n$ &0 & 0 & $n \geq 2$ & $A_{n-1}$ & $\gsu(n)$  or $\gsp(\lfloor
n/2\rfloor)$\\ 
$II$ & $\geq 1$ & 1 & 2 & none & none \\
$III$ &1 & $\geq 2$ &3 & $A_1$ & $\gsu(2)$ \\
$IV$ & $\geq 2$ & 2 & 4 & $A_2$ & $\gsu(3)$  or $\gsu(2)$\\
$I_0^*$&
$\geq 2$ & $\geq 3$ & $6$ &$D_{4}$ & $\gso(8)$ or $\gso(7)$ or $\gg_2$ \\
$I_n^*$&
2 & 3 & $n \geq 7$ & $D_{n -2}$ & $\gso(2n-4)$  or $\gso(2n -5)$ \\
$IV^*$& $\geq 3$ & 4 & 8 & $\ge_6$ & $\ge_6$  or $\gf_4$\\
$III^*$&3 & $\geq 5$ & 9 & $\ge_7$ & $\ge_7$ \\
$II^*$& $\geq 4$ & 5 & 10 & $\ge_8$ & $\ge_8$ \\
\hline
non-min &$\geq 4$ & $\geq6$ & $\geq12$ & \multicolumn{2}{c|}{ does not occur 
for susy vacua } \\ 
\hline
\end{tabular}
\end{center}
\caption[x]{\footnotesize  Table of 
codimension one
singularity types for elliptic
fibrations and associated nonabelian symmetry algebras.
In cases where the algebra is not determined uniquely by the degrees
of vanishing of $f, g$,
the precise gauge algebra is fixed by monodromy conditions that can be
identified from the form of the Weierstrass model.
}
\label{t:Kodaira}
\end{table}

A non-Higgsable gauge group factor arises on a given divisor $D$ when
all sections $f$ of ${\cal O} (-4K)$ and all sections $g$ of ${\cal
  O} (-6K)$ vanish to orders $\phi \geq 1, \gamma \geq 2$
respectively
on $D$.  In
such a situation, the  orders of vanishing  $\phi, \gamma$ force a
gauge group factor according to the Kodaira conditions in
Table~\ref{t:Kodaira}.  Note that only certain gauge groups can be
forced to appear in this way.  In particular, type $I_n$ and type
$I_n^*$ singularities with $n> 0$
cannot be forced to arise in a generic
Weierstrass model over any base.

If $f, g$ vanish to orders (4, 6) on a divisor, then there is a
``non-minimal singularity'' that cannot be resolved to give a total space
that is Calabi-Yau (and hence the data does not describe a supersymmetric
vacuum).  If $f, g$ vanish to orders $(4, 6)$ on a
codimension two locus in the base, then there is again a non-minimal
singularity.  By blowing up the codimension two locus in the base a new base arises
with a reduced degree of singularity, so that the total space of the
fibration may either be resolvable into a Calabi-Yau directly, or
after further blowups.  It is also possible to describe the structure
associated with a $(4, 6)$ vanishing on a codimension two locus in
terms of a superconformal field theory \cite{Seiberg:1996qx}; while
such field theories have been the subject of some recent work
\cite{Heckman:2013pva,
SCFT-2, SCFT-3, SCFT-4,Haghighat:2014vxa}, we do not
investigate such structure here.  In 4D models, the situation is less
clear when $f, g$ vanish to degrees $(4, 6)$ at a codimension three
locus (point).  At such points, like at $(4, 6)$ codimension two loci,
it seems that extra massless states appear in the theory
\cite{Morrison-codimension-3}; the degree of vanishing is not
sufficient at such points, however, to lead directly to a blowup of
the point -- for this we would need additional vanishing to order $(8,
12)$.  Thus, while it is possible that there is some problem or
inconsistency in models with such codimension three singularities, it
is also  plausible that such models represent perfectly acceptable
F-theory vacua.\footnote{We thank Antonella Grassi for discussions on this point.}  We do not try to resolve this question in this paper,
but we do
note some circumstances when this issue may affect some of the structures we describe
for 4D non-Higgsable clusters.

\section{Local conditions} 
\label{sec:local}

Classifying the possible non-Higgsable clusters that can arise in
F-theory compactifications to four dimensions is more difficult than
for compactifications to six dimensions, even at the level
of pure geometry.  The approach used in
\cite{clusters} to classify non-Higgsable clusters on base surfaces
incorporated a method known as the Zariski decomposition, whose 
generalization to three-dimensional bases has many complications
\cite{MR835801}.
Thus, we develop here some 
general local methods for placing constraints on the possible structure
of non-Higgsable clusters for 4D F-theory models.

For a local or global base geometry with a toric description, it is
straightforward to use the lattice of monomials dual to the lattice
containing the toric fan \cite{Fulton} to compute the orders of
vanishing of $f, g$ on any given divisor in a generic Weierstrass
model.  This method is described explicitly in \cite{toric} for base
surfaces (where it was used to analyze the set of all toric bases that support
elliptically fibered Calabi-Yau threefolds), and in \cite{Anderson-WT}
for threefold bases.  We use this approach for explicit calculations
in some specific examples in this paper, complementing the general
methods developed in this section.

In \S\ref{sec:local-derivation}, we derive local conditions that can
be used to show that $f, g$ have certain minimal orders of vanishing
on divisors on a completely general base $B$.  The results of this analysis
are summarized in \S\ref{sec:local-summary} in a succinct fashion
useful for explicit computations.

\subsection{Derivation of local conditions}
\label{sec:local-derivation}

In \cite{Anderson-WT}, a general class of 4D F-theory models were
considered where the base $B_3$ had the structure of a $\P^1$ bundle
over a complex surface $B_2$,
constructed from the projectivization of
a line bundle ${\cal L}$ over $B_2$.  
As described
in that paper, when there is a good
coordinate $z$ in an open region containing a section $\Sigma$
of the $\P^1$ bundle
(as is true, for example, when $B_2$
is toric), 
there is a series
expansion 
$f = \hat{f}_0+ \hat{f}_1z + \hat{f}_2z^2 + \cdots$ 
whose coefficients $\hat{f}_k$
restrict to sections
  of ${\cal O}_\Sigma
(-4K_{\Sigma}- (4 - k) T)$ on $\Sigma$,
and similarly $\hat{g}_k|_\Sigma \in \Gamma({\cal O}_\Sigma
(-6K_{\Sigma} - (6 - k) T)),$ where $T = c_1 ({\cal L})$ characterizes
the ``twist'' of the line bundle ${\cal L}$, and $K_\Sigma$
is the canonical class of the complex surface $\Sigma \cong B_2 \subset B_3$.
The key way that this expansion is used is in demonstrating that certain of
these coefficients must vanish upon restriction to $\Sigma$ (showing that
$f$ or $g$ must vanish to certain orders) by checking that the corresponding
line bundles on $\Sigma$ have no non-vanishing sections at all.

This characterization of vanishing conditions for $f$ and $g$ can be
made more precise and
generalized to an arbitrary divisor in a general base $B$ (of any dimension).  In the
above description, the local geometry around the divisor $\Sigma$ is
characterized by the normal line bundle, which is $N_{\Sigma} = 
N_{\Sigma/B_3} =- T$,
so the conditions on $f$ and $g$ depend only on the
local geometry and not on the global structure as a $\P^1$ bundle.  In
general, therefore, if we have a base $B$ 
containing an effective divisor $D$, we initially have
\begin{eqnarray}
f|_D & \in &\Gamma\left({\cal  O}_D\left(-4K_D + 4 N_D \right) \right)
\label{eq:f-condition-hat}\\
g|_D & \in &\Gamma\left({\cal  O}_D\left(-6K_D + 6 N_D \right)\right) \,,
\label{eq:g-condition-hat}
\end{eqnarray}
where 
$K_D$ and
$N_D = N_{D/B}$ are the canonical and normal line bundles for $D\subset B$,
and we have used the adjunction formula, which tells us that $-K_B|_D=-K_D+N_D$.

There are exact sequences that help to measure the vanishing of $f$ and
$g$ along $D$:
\begin{eqnarray}
0 &\to \mathcal{O}(-4K_B-D) \to \mathcal{O}(-4K_B) \to \mathcal{O}_D(-4K_D+4N_D) \to 0\\
0 &\to \mathcal{O}(-6K_B-D) \to \mathcal{O}(-6K_B) \to \mathcal{O}_D(-6K_D+6N_D) \to 0 .
\end{eqnarray}
Thanks to these sequences, if $f|_D$ vanishes 
then
we can write $f=\hat{f}_1z$ 
with $\hat{f}_1$ a section of 
$\mathcal{O}(-4K_B-D)$.
Similarly, if
 $g|_D$ vanishes then we can write
$g=\hat{g}_1z$ with
 $\hat{g}_1$ a section of $\mathcal{O}(-6K_B-D)$.

We can continue, and try to detect if $f$ or $g$ vanishes to order $2$.
For this purpose, we use the pair of exact sequences
\begin{eqnarray}
0 &\to \mathcal{O}(-4K_B-2D) \to \mathcal{O}(-4K_B-D) \to \mathcal{O}_D(-4K_D+3N_D) \to 0\\
0 &\to \mathcal{O}(-6K_B-2D) \to \mathcal{O}(-6K_B-D) \to \mathcal{O}_D(-6K_D+5N_D) \to 0 .
\end{eqnarray}
To understand these sequences it is helpful to recall that $\mathcal{O}_D(-D)$
is an alternate way of writing the line bundle $\mathcal{O}_D(-N_D)$, and we
have used this equivalence in the exact sequence.  Note that $\hat{f}_1$ or $\hat{g}_1$,
when they exist, are sections of the middle term in the exact sequence; we
restrict them to $D$, and if  one of them is zero, then we will be able 
to write $\hat{f}_1=\hat{f}_2z$ (or $\hat{g}_1=\hat{g}_2z$), i.e., $f=\hat{f}_2z^2$ (or $g=\hat{g}_2z^2$).
This happens if and only if $f$ (respectively $g$) vanishes to order at least
$2$ along $D$.

We now see the general pattern:  if $f$ vanishes to order at least $k$
then we can write $f=\hat{f}_kz^k$ with $\hat{f}_k$ a section of
$\mathcal{O}(-4K_B-kD)$ and restrict $\hat{f}_k$ to $D$.  The corresponding
exact sequence is 
\begin{equation}
0 \to \mathcal{O}(-4K_B-(k+1)D) \to \mathcal{O}(-4K_B-kD) \to \mathcal{O}(-4K_D+(4-k)N_D) \to 0 .
\end{equation}
The restriction vanishes if and only if $f$ vanishes to order at least
$k+1$ along $D$, and in that case we can write $\hat{f}_k=\hat{f}_{k+1}z$ so that
$f=\hat{f}_{k+1}z^{k+1}$.

Similarly, if $g$ vanishes to order at least $k$
then we can write $g=\hat{g}_kz^k$ with $\hat{g}_k$ a section of
$\mathcal{O}(-6K_B-kD)$ and restrict $\hat{g}_k$ to $D$.  The corresponding
exact sequence is 
\begin{equation}
0 \to \mathcal{O}(-6K_B-(k+1)D) \to \mathcal{O}(-6K_B-kD) \to \mathcal{O}(-6K_D+(6-k)N_D) \to 0 .
\end{equation}
The restriction vanishes if and only if $g$ vanishes to order at least
$k+1$ along $D$, and in that case we can write $\hat{g}_k=\hat{g}_{k+1}z$ so that
$g=\hat{g}_{k+1}z^{k+1}$.

More generally,
given a set of effective
divisors $D_a$ in $B$ together with the information that
$f$ vanishes on $D_a$ to order at least $\phi_a$ and $g$ vanishes
on $D_a$ to order at least $\gamma_a$, then we can write
\begin{eqnarray}
f &= {f}^{[a]} \prod_{b\ne a} z_b^{\phi_b} \\
g &= {g}^{[a]} \prod_{b\ne a} z_b^{\gamma_b} 
\end{eqnarray}
where $z_b$ is a local coordinate vanishing on $D_b$.

We now go through the same reasoning with analogous exact sequences,
starting from $-4K_B-\sum_{b\ne a}\phi_bD_b$ instead of $-4K_B$,
and 
$-6K_B-\sum_{b\ne a}\gamma_bD_b$ instead of $-6K_B$, to determine the orders of vanishing
of ${f}^{[a]}$ and ${g}^{[a]}$. If the order of vanishing is
at least $k$, then the restricted leading coeffient lies in
\begin{equation}
{f}^{[a]}_k|_{D_a} \in \Gamma(\mathcal{O}_{D_a}(-4K^{(a)}+(4-k)N^{(a)} -
\sum_{b\ne a} \phi_b C_{ab}))
\label{eq:f-condition-1}
\end{equation}
or
\begin{equation}
{g}^{[a]}_k|_{D_a} \in \Gamma(\mathcal{O}_{D_a}(-6K^{(a)}+(6-k)N^{(a)} -
\sum_{b\ne a} \gamma_b C_{ab})) ,
\label{eq:g-condition-1}
\end{equation}
where  $C_{a b}= D_a \cap D_b$, considered as a curve in $D_a$.  The
terms proportional to $C_{a b}$ arise because the vanishing of $f, g$
around $D_b$ appear on $D_a$ as additional vanishings on the
curves $C_{ab}$.   

The properties of the bundles in (\ref{eq:f-condition-1}) and (\ref{eq:g-condition-1}) can
be used to determine the minimal possible orders of vanishing $\phi_a,
\gamma_a$ of $f, g$ on each divisor $D_a$ in a self-consistent
fashion.  By using the fact that {\it e.g.}  $f^{[a]}_k|_{D_a}$ must vanish
if the line bundle of which it is a section corresponds to a
non-effective divisor on $D_a$, (\ref{eq:f-condition-1}) and
(\ref{eq:g-condition-1}) specify a collection of bundles which can be
checked for the existence of non-zero sections, and if those sections
are absent, the corresponding leading coeffients must vanish (i.e.,
the order of vanishing will be greater than might have been expected).

For the monodromy conditions
associated with each divisor, and to identify the complete (geometric)
matter content, we need to consider these coefficients more generally
as sections of (\ref{eq:f-condition-hat}, \ref{eq:g-condition-hat})
that vanish to orders $\phi_b, \gamma_b$ on $C_{a b}$, and we use
$\hat{f}^{[a]}_k|_{D_a}, \hat{g}^{[a]}_k|_{D_a}$ in such situations. 

\subsection{Summary of local conditions}
\label{sec:local-summary}

We summarize here  the constraints derived in the previous
section and define some notation that will be useful for explicit
calculations.  
For a compactification of F-theory on a base $B$, for each
effective divisor $D_a$ in $B$ we define corresponding families of
divisors
\begin{eqnarray}
F_k^{(a)} & = & -4K^{(a)}+ (4 - k) N^{(a)} -\sum_{b\ne a} \phi_b
C_{ab}\label{eq:f-divisors}\\
G_k^{(a)} & = & -6K^{(a)}+ (6 - k) N^{(a)} -\sum_{b\ne a} \gamma_b
C_{ab}
\label{eq:g-divisors}
\,.
\end{eqnarray}
Here, as above, $- K^{(a)}, N^{(a)}$ are the divisors associated with the
anti-canonical and normal line bundles to $D_a$, $\phi_a, \gamma_a$ are
the orders of vanishing of $f, g$ on $D_a$, and $C_{ab}$
is the curve $D_a \cap D_b$
considered as a divisor class  on $D_a$.

When there is no effective divisor in any of the divisor classes
$F_j^{(a)}$ for $j = 0, 1, \ldots, k -1$, then $f$ must vanish to at
least order $k$ on $D_a$.
Similarly,
when there is no effective divisor in any of the divisor classes
$G_j^{(a)}$ for $j = 0, 1, \ldots, k -1$, then $g$ must vanish to at
least order $k$ on $D_a$.
This determines a set of conditions on the vanishing orders of $f,
g$ on different divisors in the base
that must be satisfied in a self-consistent fashion. 
Note that these conditions determine a \emph{minimum} order of
vanishing of $f, g$ on each divisor through the structure of the local
geometry.  We have not ruled out the possibility that
further nonlocal structure may force $f, g$ to vanish to
higher orders in some circumstances.

In the following sections we use 
the divisors specified in (\ref{eq:f-divisors}) and                           
(\ref{eq:g-divisors}) to analyze various
situations in which the vanishing of $f, g$ to particular orders
guarantees the existence of non-Higgsable clusters of different kinds
in Calabi-Yau threefolds and fourfolds corresponding to F-theory
compactifications to 6D and 4D respectively.

In general the  restriction of the
leading non-vanishing term in an expansion of $f$ around the divisor $D_a$
can be  described as a section of the line bundle
over $D_a$ associated with $F_k$
\begin{equation}
 f_k^{(a)}= f_k^{[a]}|_{D_a} \in \Gamma ({\cal O}_{D_a} (F_k)) \,,
\;\;\;\;\;k=\phi_a
\label{eq:f-condition}
\end{equation}
and the restricted leading term in $g$ can similarly be described as a section of
\begin{equation}
 g_k^{(a)}= g_k^{[a]}|_{D_a} \in \Gamma ({\cal O}_{D_a} (G_k))  \,,
\;\;\;\;\;k=\gamma_a\,.
\label{eq:g-condition}
\end{equation}
Note that for a general base
these equations are only meaningful  
for the first non-vanishing term in each of $f$ and $g$, though in
special cases such as toric bases where there are good global
coordinates, these expressions are valid for all $k$.

As discussed above, when determining monodromy conditions and matter
content, it is useful to consider the leading terms $\hat{f}^{(a)}_k
=\hat{f}^{[a]}_k|_{D_a}, \hat{g}^{(g)}= \hat{g}^{[a]}_k|_{D_a}$ as
sections of the line bundles $\Gamma ({\cal O}_{D_a}(\hat{F}_k))$,
$\Gamma ({\cal O}_{D_a}(\hat{G}_k))$, with $\hat{F}_k = -4K^{(a)}+ (4
- k) N^{(a)}$, $\hat{G}_k = -6K^{(a)}+ (6 - k) N^{(a)}$.

\subsection{Additional constraints}
\label{sec:additional-constraints}

We conclude this section by briefly mentioning a further local constraint
that is not used directly in the analysis of this paper, but which may
be useful in further analyzing the set of possible local divisor
configurations and associated non-Higgsable clusters in general
F-theory models.

In addition to the constraints described in the preceding sections, we
have the geometric constraint
\begin{equation}
N^{(b)} \cdot C_{b a} = C_{a b} \cdot C_{a b} \,.
\label{eq:geometric-constraint}
\end{equation}
As can be inferred from the notation, the intersection on the left is
carried out within $D_b$, while that on the right is in $D_a$.

This constraint follows from a general fact about intersection theory:
if $D_1$, $D_2$, and $D_3$ are three divisors, then $D_1\cdot D_2 \cdot D_3$
can be computed as an intersection of two divisors on $D_3$, namely,
the intersection of the divisors $D_1|_{D_3}$ and $D_2|_{D_3}$.
Permuting the $D_i$'s gives multiple ways to compute the same intersection
property.  To apply this to derive (\ref{eq:geometric-constraint}), 
we consider the intersection product $D_b\cdot D_b\cdot D_a$.
On the one hand, this can be evaluated on $D_b$ as the intersection of
$D_b|_{D_b}=N_D $ with $D_a|_{D_b} = C_{b a}$.  On the other hand,
the same triple intersection can be evaluated on $D_a$ as the
intersection of $D_b|_{D_a}$ with $D_b|{D_a}$, i.e., as
$C_{a b}\cdot C_{a b}$.

In the toric situation, the relation (\ref{eq:geometric-constraint})
follows directly from the structure of the  fan for a toric
threefold.  Assuming that the geometry is smooth, and
taking a choice of coordinates where $D_a, D_b$  are associated with
rays $v_a =(0, 0, 1),  v_b =(0, 1, 0)$,
and there are 3D cones connecting these two rays to the rays $v_c =(1, 0,
0)$ and $v_d =(-1, y, z)$, we see that both sides of
(\ref{eq:geometric-constraint}) are identified with the value
$ -y$. In the left-hand side, we have $N^{(b)} = -y C_{bd}$ and $C_{bd}
\cdot C_{ba} = 1$, so $N^{(b)} \cdot C_{ba} = -y$, and on the
right-hand side we have the same result since $C_{ab}$ is a curve of
self-intersection $-y$ from the fact that on projection to the plane
$z = 0$, $y v_b = v_c + v_d$.

\section{Warm-up: 6D non-Higgsable clusters} 
\label{sec:6D}

As an illustration of how the constraints derived in the previous section
can be used to characterize non-Higgsable
clusters, we begin as a warm-up exercise with the 6D case.
A complete classification of non-Higgsable clusters for 6D F-theory
compactifications was given in \cite{clusters}.  Here we show how
these results can  be reproduced easily using  the constraint equations
 derived in the previous section.

\subsection{Constraints on individual curves}

In six dimensions, 
we are concerned with
Calabi-Yau threefolds that are elliptically fibered over a complex base
surface $B_2$.  In this situation, the divisors that support
codimension one singularities of the elliptic fibration associated
with gauge group factors are curves, and codimension two singularities
are associated with points.  This simplifies the analysis
significantly, since all points on a curve represent the same homology class,
so we can represent all divisors on a curve simply as an integer in
$\Z$.
Specializing the discussion of the previous section to the case of
a base of dimension $2$, and denoting the divisors by $C_a$ since they are
now curves on the base, we find
\begin{eqnarray}
F^{(a)}_k& = &-4K^{(a)} + (4 - k) N^{(a)} - \sum_{b}
\phi_b Z_{a
  b} \,,
\label{eq:f-condition-surface}\\
G^{(a)}_k  &= &
-6K^{(a)} + (6 - k) N^{(a)} - \sum_{b}
\gamma_b Z_{a
  b} \label{eq:g-condition-surface} \,.
\end{eqnarray}
Here, as before, $K^{(a)}$ is the canonical class of $C_a$, $N^{(a)}$
is the class of the normal bundle, while $Z_{a b}$ is the intersection
of $C_a$ with $C_b$, considered as a zero-cycle on $C_a$.
(When $C_a$ is a rational curve, the only thing that matters
about this zero-cycle is its degree $p_{a b}=\deg Z_{a b}$,
which is the intersection number
$C_{a}\cdot C_b$).

We begin by noting that if the anti-canonical class $- K_C = -
K^{(a)}$ is not
effective
({\it i.e.} a nonnegative integer class) for a given curve $C = C_a$, then
(dropping the superscript $(a)$ henceforth on $f_k, g_k$, which we
take to be assumed for any given curve $C$)
$F_k, G_k$ could not be effective and
$f_k,g_k$
would vanish for all
$k$ unless $N_C$ were effective, in which case no $f_k, g_k$ could be
non-vanishing unless the same were true of $f_0, g_0$.  
Similarly, if $-K_C = 0$ then either all $f_k, g_k$ can be
nonvanishing or none can.
Since $-K_C = 2-2g$ on an irreducible curve of genus $g$,
this leads us
to the conclusion that there cannot be a non-Higgsable cluster on any
curve of higher genus; this was shown from a different point of view
in \cite{clusters}.

We assume then that all irreducible
curves $C_a$ supporting a non-Higgsable
cluster are rational curves ({\it i.e.}, equivalent to $\P^1$). We
have then $- K^{(a)}= 2$, and $N^{(a)}= C_a \cdot C_a$ is the self-intersection of
$C_a$.

Consider for example the case where
$C_a$ is a curve of self-intersection $-2$.  In this case, $\deg F_k
= 2k - \phi$, $\deg G_k =2k - \gamma$, where $\phi=
\sum_{b\neq a}\phi_b p_{a b}$.  If $\phi = 0$, so there is no forced
vanishing of $f$ on any curves that intersect $C_a$, then 
$\deg F_0 = \deg G_0 = 0$.  We then have
$f_0 \in
{\cal O} (0) =\C$, and similarly for $g_0$, so there is no forced
vanishing of $f, g$ on $C_a$.

Now consider the case of a curve of self-intersection $N^{(a)}= -3$.
In this case, 
\begin{eqnarray}
\deg F_k &  = &  -4 +3k - \phi\\
\deg G_k & = & -6 +3k - \gamma \,.
\end{eqnarray}
It follows that $f_0, f_1, g_0, g_1$ must all vanish, so the curve
$C_a$ must support a Kodaira type IV (2, 2, 4) singularity.
Furthermore, $g_2 = \hat{g}_2 \in {\cal O} (0) =\C$ (assuming $\phi = 0$), which
satisfies the monodromy condition so that the associated gauge group
is $SU(3)$.  This analysis is essentially equivalent to the Zariski
decomposition method used in \cite{clusters}, in which $- K$ is
decomposed over the rationals, so that for a self-intersection $-3$
curve $C$, with $- K \cdot C = -1$, $- K = C/3 + X$ with $X$ ($\Q$-)effective
(actually nef) from which it follows that $-4K$ contains two factors
of $C$ as irreducible components, as does $-6K$.  The method of
analysis used here, however, generalizes more readily to
four-dimensional F-theory compactifications than the Zariski approach.

\begin{table}
\begin{center}
\begin{tabular}{|r |c |c |l |}
\hline
$C \cdot C$ & Divisors $F_k$ & Divisors $G_k$ & singularity type \\
\hline
$-1$ & $\deg F_k=4 + k - \phi$ &
$\deg G_k = 6 + k - \gamma$ &
$(0, 0, 0)$\\
$-2$ & $\deg F_k=2k - \phi$ &
$\deg G_k = 2k -\gamma$ &
$(0, 0, 0)$\\
$-3$ & $\deg F_{2 + n}=2 + 3n - \phi$ &
$\deg G_{2 + n} = 3n -\gamma$ &
$(2, 2, 4) \Rightarrow IV (\gsu_3)$\\
$-4$ & $\deg F_{2 + n}=4n - \phi$ &
$\deg G_{3 + n} = 4n -\gamma$ &
$(2,  3, 6) \Rightarrow I_0^* (\gso_8)$\\
$-5$ & $\deg F_{3 + n}=3 + 5n - \phi$ &
$\deg G_{4 + n} = 2 + 5n -\gamma$ &
$(3, 4, 8) \Rightarrow  IV^* (\gf_4) $\\
$-6$ & $\deg F_4= 2 + 6n - \phi$ &
$\deg G_{4 + n} =  6n -\gamma$ &
$(3, 4, 8) \Rightarrow  IV^* (\ge_6)$\\
$-7$ & $\deg F_{3 + n}=  1+ 7n - \phi$ &
$\deg G_{5 + n} =  5 + 7n -\gamma$ &
$(3,  5, 9) \Rightarrow  III^* (\ge_7)$\\ 
$-8$ & $\deg F_{3 + n} =   8n - \phi$ &
$\deg G_{5 + n} =   4 + 8n -\gamma$ &
$(3,  5, 9) \Rightarrow  III^* (\ge_7)$\\ 
$-9/10/11$ & $\deg F_{ 4}  =   8- \phi$ &
$\deg G_{5} =   3/2/1  -\gamma$ &
$( 4, 5, 10) \Rightarrow  II^* (\ge_8)$\\ 
$- 12$ & $\deg F_{4}=  8- \phi$ &
$\deg G_{5} =  -\gamma$ &
$( 4, 5, 10) \Rightarrow  II^* (\ge_8)$\\ 
\hline
\end{tabular}
\end{center}
\caption[x]{\footnotesize
Table of degrees of divisors $F_k, G_k$ and resulting non-Higgsable
singularity types on single rational curves $C$ of self-intersection $-1$
through $-12$.
}
\label{t:6D-table}
\end{table}

Systematically applying these methods for any irreducible rational
curve $C_a$ of given self-intersection, the divisors $F_k, G_k$ are
easily computed and determine the orders of vanishing of $f, g$ over
the curve $C_a$, as tabulated in Table~\ref{t:6D-table}.  From the
data in this table, we can determine many features of the gauge groups
and matter that arise at generic points in complex structure moduli
space for bases that contain one or more intersecting curves of
negative self-intersection.  In particular, we can determine the
precise minimal gauge group, including effects of monodromy; we can
ascertain the generic matter content; we can identify cases where
there is a (4, 6) singularity at a point; and we can classify
non-Higgsable clusters containing multiple gauge group factors.  We
describe a few details of each of these aspects in the following
subsections.

\subsection{Monodromy} 

In the cases of Kodaira singularities of types $IV, I_0^*$, and
$IV^{*}$, the gauge group of the low-energy theory depends upon an
additional monodromy condition; the Dynkin diagram describing the set
of cycles produced when a codimension one singularity is resolved can
be mapped to itself non-trivially under a closed path in the relevant
divisor that goes around a codimension two singularity.  The details
of how the gauge group is determined in this case are worked out in
\cite{Bershadsky-all, Grassi-Morrison-2}; when considering the generic
structure in the moduli space as is relevant for non-Higgsable
clusters, the monodromy condition can be read off directly from the
form of the leading terms in $f, g$ in the expansion around the
divisor.  These monodromy conditions on monomials for non-Higgsable
clusters are described briefly in \S9 of \cite{Anderson-WT},
and  analyzed and explained further in Appendix A.  These
monodromy conditions are valid for F-theory compactifications in any
dimension below eight, and will also be used in analyzing
compactifications to 4D in later sections.

For type $IV$ and $IV^{*}$, the monodromy is determined by the leading
coefficient in $g$.  For a type $IV$ codimension one singularity, if
$\hat{g}_2$ is a perfect square, then there is no nontrivial monodromy and
the gauge algebra is $\gsu_3$; otherwise it is $\gsu_2$.  
For $\hat{g}_2$ to generically be a perfect square, every section
of ${\cal O}_D(-6K_D+4N_{D/B})$ that comes from the restriction of
a section of ${\cal O}_B(-6K_B-2D)$ must be the square of a section
of ${\cal O}_D(-3K_D+2N_{D/B})$.  (In particular, it's not hard to see
that the space of these sections can only be one-dimensional.)
This is clearly the case for a $-3$ curve that does not
intersect any other curves where $g$ vanishes ($\gamma = 0$), 
where $\hat{g}_2 \in{\cal O} (0)$
so the
non-Higgsable gauge group there has an algebra $\gsu_3$.  Similarly,
for a type $IV^{*}$ singularity the gauge algebra is $\ge_6$ (no
monodromy) if $\hat{g}_4$ is a perfect square, and $\gf_4$ otherwise.  This
allows us to immediately read off the $\gf_4$ and $\ge_6$ gauge
algebras of the non-Higgsable cluster over curves of self-intersection
$-5, -6$ respectively.   For a
non-Higgsable type $I_0^*$ singularity, the
gauge algebra is $\gso_8$ (no monodromy) only when 
$\hat{f}_2, \hat{g}_3$ are both
in one-dimensional spaces of sections $\Gamma
({\cal O} (2 X)),  \Gamma ({\cal O} (3
X))$, and are proportional to second and third powers of some section
$u$ in the one-dimensional space of sections
$\Gamma({\cal O} (X))$, where
$X = -K_D+N_{D/B}$. 
 In this case,
the cubic $x^{3}+ \hat{f}_2 x + \hat{g}_3$ can be algebraically
factorized to a product $(x - A) (x - B) (x - C)$ for generic choices of $f, g$.  This condition is clearly
satisfied for the non-Higgsable cluster over a $-4$ curve, where $X = 0$.  The
remaining monodromy condition is that for a non-Higgsable type $I_0^*$ singularity,
we have a gauge algebra $\gso_7$
when  $x^{3}+ \hat{f}_2 x + \hat{g}_3$
factorizes into the product of a quadratic times a
linear term for generic $f, g$.  When $f$ and $g$ are generic
so that maximal Higgsing has been done, this occurs only when $\hat{g}_3$ 
vanishes identically and $\hat{f}_2$ is not a perfect square; in all other
cases the gauge algebra (of generic models) is $\gg_2$.  The $\gso_7$ condition does not
occur for any of the non-Higgsable clusters over a single curve, but
does occur over a combination of curves $-2, -3, -2$ as discussed
below.

\subsection{Matter} 

Matter can arise in F-theory constructions either from nonlocal
structure, associated  in 6D compactifications with the genus of the
divisor on which a gauge group is supported, or from local structure
associated with codimension two singularities.  For gauge groups
without monodromy, the presence of matter can be identified when there
are codimension two loci on the gauge group divisor where the Kodaira
singularity type is enhanced.  A specific example of this can be seen
for a $-7$ curve, where $F_3=1 - \phi$
and $\hat{f}_3 \in {\cal O} (1)$, so generically
there is a point where $\hat{f}_3$ vanishes and the singularity type becomes
$II^{*}$ (4, 5, 10).  This corresponds to the appearance of a
half-hypermultiplet in the ${\bf 56}$ representation.
For the other single non-Higgsable gauge group
factors without monodromy, such as $\gsu_3$ on a $-3$ curve, there are
no points where the Kodaira singularity type is enhanced, since {\it
  e.g.} $g_2 \in \Gamma ({\cal O} (- \gamma)) =\C$ when $\gamma = 0$.  For a
$-5$ curve, there are generically two points where $g_4$ vanishes.
These, however, are the points around which there is monodromy; a
double cover of $\P^1$ with two branch points is again a $\P^1$, and
there is again no matter in this case
(see {\it e.g.}  \cite{Katz-Morrison-Plesser, Grassi-Morrison}).

\subsection{Superconformal fixed points} 

Analogous to the appearance of matter, at loci on a $II^{*}$ curve
where $g_5$ vanishes, there is a $(4, 6, 12)$ vanishing of $(f, g,
\Delta)$.  Such points correspond to theories where gravity is coupled
to a superconformal field theory \cite{Seiberg:1996qx}; these are
branch points in the moduli space that are associated with tensionless
string transitions \cite{Ganor-Hanany,Seiberg-Witten}.  By blowing up
the $(4, 6)$ point in the base, one enters a different branch of the
moduli space where the 6D theory has an extra tensor multiplet
\cite{Morrison-Vafa-II,Heckman:2013pva}. 
Such tensionless string transitions unify the space of all 6D
F-theory compactifications  into a single connected space
\cite{Grassi, KMT-II, mt-sections}.

\subsection{Clusters with multiple factors} 
\label{sec:multiple-factors}

From Table~\ref{t:6D-table} we can also determine the set of possible
non-Higgsable clusters containing multiple intersecting curves of
negative self-intersection, reproducing the results of
\cite{clusters}.  Restricting attention to bases that do not include
$(4, 6)$ points, clearly there cannot be an intersection between two
curves where the orders of vanishing of $f, g$ add to $(4, 6)$ or
more.  This rules out any intersection between two curves each of
self-intersection $-4$ or below.  Even for two intersecting $-3$
curves, since $g$ vanishes on each to order at least two, we must have
for each $\gamma \geq 2$, which implies $g_2 = 0$ on each, so there is
a $(4, 6)$ point at the intersection.  Any intersection between a $-3$
curve and a curve of self-intersection $-4$ or below is even worse.
So the only intersections that we need to consider for curves of
self-intersection $- 2$ or below are between $-2, -2$ or $-2, -3$
curves.  Any combination of $-2$ curves alone cannot give rise to a
non-Higgsable factor since we can have $ f_0, g_0 \in
\Gamma ({\cal O} (0)) =\C$
on each $-2$
curve.  Considering configurations with $-3$ and $-2$ curves, it is
easy to check that the only nontrivial combinations giving
non-Higgsable clusters are the $-2, -3$ and $-2, -2, -3$ combinations
giving $\gsu_2\oplus \gg_2$ algebras and the $-2, -3, -2$ cluster that
gives $\gsu_2 \oplus \gso_7 \oplus\gsu_2$, as found in
\cite{clusters}.  For example, on a $-2$ curve $C$ that intersects a
$-3$ curve $D$, we must have $\phi_C, \gamma_C \geq 2$, which implies
${\rm ord}_C f, {\rm ord}_C g \geq 1$, which pushes $\gamma_D \geq 1$,
so $g^{D}_2 = 0$, so $\gamma_C \geq 3$, and we have then type $I_0^*$
and $III$ singularities on the $-3, -2$ curves respectively, with
generic monodromy on the $I_0^*$ since $g_3^{D} \in \Gamma ({\cal O} (1))$ is
nonzero and is not a constant.  This reproduces the $\gg_2 \oplus
\gsu_2$ gauge algebra found in \cite{clusters} for the $-3, -2$
non-Higgsable cluster.  The story is similar the other cases.  Note
that for the $-2, -3, -2$ cluster, on the $-3$ curve we have $\gamma =
4$, so $g_3 = 0$, while $\phi = 2$, so $f_2 \in \Gamma ({\cal O} (0))$ and
$\hat{f}_2 \in \Gamma ({\cal O} (2))$, where $\hat{f}_2$ has two distinct roots
(corresponding to the points of intersection with the two $-2$ curves)
and is not a perfect square, so $x^{3}- \hat{f}_2 x$ cannot factorize
completely and the monodromy is $\gso_7$.  Further analysis of this
type can be used to confirm the various combinations of non-Higgsable
clusters that can be connected by $-1$ curves as enumerated in
\cite{clusters}.

\section{4D non-Higgsable clusters with single gauge group factors}  
\label{sec:4D-single}

We now turn to F-theory compactifications to four dimensions, which
involve compactification on Calabi-Yau fourfolds that are elliptically
fibered over a threefold base $B_3$.  While the story is in some ways
parallel to that of six dimensions, there are a number of additional
complications for four-dimensional theories, and the set of possible
non-Higgsable clusters seems to be substantially richer than for 6D
models.  One issue that makes a general analysis of non-Higgsable
clusters in threefold bases more complicated than in twofold bases 
is the wide range of possible surfaces that can
arise as divisors in the threefold base.  In the case of Calabi-Yau
threefolds, as discussed in the previous section, curves in the
twofold base are classified by genus, and the only curve topology
that can support a non-Higgsable cluster is a $\P^1$.  In base
threefolds, on the other hand, a vast range of surfaces can be
realized as divisors.  Restricting to toric surfaces alone, each of
the 61,539 toric surfaces enumerated in \cite{toric} can arise as a
codimension one divisor in a threefold base that supports an elliptic
fibration (in the simplest case by taking a product with $\P^1$).
Hundreds of distinct choices of these divisor geometries can support
non-Higgsable clusters \cite{Halverson-Taylor}.  A complete analysis
of all algebraic surfaces that can act as divisors supporting
non-Higgsable clusters represents a substantial project for future
investigation.

The simplest non-Higgsable gauge groups are single nonabelian factors.
We describe the possible single factor groups in
\S\ref{sec:single-factors}.  In \S\ref{sec:single-matter} we describe
the possible appearance (at the level of geometry) of matter localized
on curves in the threefold base, which presents richer possibilities
than in 6D.

\subsection{Possible single gauge factors} 
\label{sec:single-factors}

The set of possible isolated simple gauge algebras for 4D models is
basically the same as for 6D, with the additional possibilities of
$\gsu_2$ and  $\gso_7$.
The only possibilities  from the Kodaira table
that are ruled out, in fact, are those where the order of vanishing of
$\Delta$ exceeds that determined by $f, g$: 
\begin{equation}
 {\rm ord} (\Delta) > {\rm max} (3\,{\rm ord} (f), 2\,{\rm ord} (g)) \,.
\end{equation}
For this to occur, we would need to have a cancellation between the
leading terms in $\Delta = 4f^{3}+ 27g^2$.  But such a cancellation
cannot occur between generic sections, since we can always multiply
$f$ and $g$ by different constant complex factors and preserve the
section property while eliminating the cancellation of leading terms.

Thus, the only possible nonabelian gauge algebra components that can
be realized are
\begin{equation}
 \gsu_2, \gsu_3, \gg_2, \gso_7,  \gso_8, \gf_4, \ge_6, \ge_7, \ge_8 \,.
\end{equation}

We can  identify explicit examples in which each of these gauge algebras
is realized in a non-Higgsable cluster.  The
simplest set of examples corresponds to the case where the divisor
supporting the gauge group is the surface $S = \P^2$.  In this case,
similar to the 6D case where non-Higgsable gauge groups arise on
rational curves $\P^1$, all divisors in the surface $S$ are linearly
equivalent to a multiple of the hyperplane class $H$, so the relevant line bundles
can be classified by a single integer.  In particular, the normal
bundle can be chosen so that $N=  -nH$ for any integer $n$.  This
local condition can be realized explicitly in the context of a global model by taking a compact
base $B_3 = \tilde{\F}_n$ that is a $\P^1$ bundle over $\P^2$ formed by
projectivization of the line bundle ${\cal O} ( -nH)$.  Such geometries
were described previously in \cite{Klemm-Yau, Berglund-Mayr,
Grimm-WT,  Anderson-WT}.  In this case, the
divisors (\ref{eq:f-divisors})
and (\ref{eq:g-divisors}) become
\begin{eqnarray}
F_k &   = &(12 -(4 - k) n)H\label{eq:fg-p2}\\
G_k & = &  (18  -(6 - k) n)H \,, \nonumber
\end{eqnarray}
where we assume $\phi = \gamma = 0$ so that the divisor $S$ does not
intersect other divisors on which $f, g$ must vanish, in accord with
the assumption that this is an isolated single-factor non-Higgsable
cluster.  It is straightforward to read off the Kodaira singularity
types on $S$ associated with different values of $n$.  For example,
for $n= 4,$ $F_k = (4k-4)H, G_k = (4k -6)H,$ so
$(f, g)$ vanish to degrees $(1, 2)$, corresponding to a type $III$
codimension one singularity supporting an $\gsu_2$ gauge algebra.
Similarly, for $n= 5, 6, \ldots, 12$, the associated gauge algebras
are $\gg_2, \gso_8, \gf_4, \gf_4,  \ge_6, \ge_7, \ge_7, \ge_7,$ and
for $n= 18$ the gauge algebra is $\ge_8$.  For $13 \leq n\leq 17$,
there is a codimension two $(4, 6)$ locus associated with a curve on
$S$, analogous to the $(4, 6)$ points on curves of self-intersection
$-9, -10, -11$ in the 6D case.  This gives explicit examples of all
the single-factor non-Higgsable gauge group possibilities other than
$\gsu_3$ and $\gso_7$.

While $\gsu_3$ cannot be realized as a single-factor non-Higgsable
cluster on a divisor $\P^2$ that does not intersect other divisors on
which $f, g$ vanish, $\gsu_3$ can be realized on divisors that realize
other types of surfaces. 
Several explicit examples were given in \cite{Anderson-WT, ghst};
in one case, if $S =\F_0 = \P^1 \times \P^1$,
which has $- K = 2 S + 2F$ where $S, F$ are the curves associated with
the two $\P^1$ factors, we can take $N= -3 S -3F$, and we have
\begin{eqnarray}
F_k & = &  (3k-4) (S + F)\\
G_k & = &  (3k-6) (S + F) \,.
\end{eqnarray}
It follows that $f_0 = f_1 = g_0 = g_1$, and $g_2, \hat{g}_2 \in {\cal
  O} (0)$, so we have a type $IV$ singularity with no monodromy, and
the gauge algebra is $\gsu_3$.
This geometry can be realized in the context of a $\P^1$ bundle over
$\F_0$, produced from projectivization of a line bundle ${\cal O} (3 S
+ 3F)$.

It may seem na\"{\i}vely that it should not be possible to realize $\gso_7$
as a single-factor non-Higgsable cluster on any divisor that does not
intersect other divisors on which $f, g$ vanish.  In this case we must
have
$ f_2 \in {\cal O} (2 X), g_3 \in {\cal O} (3 X) $, where $X = -2K +
2N$, and one might think that if $2 X$ is effective then so is $3 X$.  
This is not true, however, for surfaces in threefold bases,
illustrating one of the subtle aspects of generalizing the Zariski
decomposition to threefolds.
The issue is that $-2K$ may be an effective divisor, while $-3K$ may
only be in the effective cone
({\it e.g.} an effective $\Q$-divisor) without a realization as an integer
linear combination of irreducible algebraic hypersurfaces.

An explicit example of an isolated $\gso_7$ non-Higgsable cluster that
illustrates this phenomenon can be constructed as follows.  Consider
a set of toric divisors $B, D, F, H$ on $\F_0$ that each have
self-intersection zero and intersect cyclically with intersections $H \cdot B=B
\cdot D = D \cdot F = F \cdot H = 1$.  Now blow the surface up at
these four intersection points giving exceptional divisors $A, C, E,
G$.  We then have a toric surface $S$ with a set of toric divisors $A$-$H$
having
self-intersections $(-1, -2, -1, -2, -1, -2, -1, -2)$.  The
anti-canonical class is
\begin{equation}
- K = A + B + C + D + E + F + G + H
\end{equation}
and there are two equivalence relations (from the Stanley-Reisner
ideal)
\begin{equation}
A + B + C \sim E + F + G, \;\;\;\;\;
C + D + E \sim G + H + A \,.
\end{equation}
Now, we consider embedding this surface into a threefold with normal
bundle
\begin{equation}
N= - C -2 D -4E -3F -3G - H\,.
\end{equation}
We can then write
\begin{equation}
-4K + 2N = B + D + F + H\,,
\end{equation}
which is clearly effective, so $f_2$ is generically nonzero.  In this
geometry, however, the divisor $-6K + 3N = 3A + 3B + 3H -3E$, while in
integer homology cannot be written as a sum of effective irreducible
divisors with nonnegative integer coefficients.  Thus, in this case 
$\hat{g}_3$ vanishes, $\hat{f}_2$ is not a perfect square, and
we
have an isolated $\gso_7$ non-Higgsable cluster.  This local geometry
can be realized globally by simply considering a $\P^1$ bundle over
the surface $S$ with an appropriate twist $T = -N$.  An explicit
computation of the orders of vanishing of $f, g$ on the divisors in
this threefold base using toric methods
confirms the presence of a non-Higgsable $\gso_7$
\cite{Anderson-WT}.

\subsection{Matter} 
\label{sec:single-matter}

While in 6D the only single gauge factor that can have associated
non-Higgsable
matter is $\ge_7$,  4D constructions
can provide a much richer range of matter associated with isolated
non-Higgsable gauge factors.  The  primary reason for this difference
is that while the possible classes of zero-cycles
that could
support a codimension two matter locus for a 6D theory
are labeled by a single integer (the degree),
in 4D, the codimension two matter is supported on curves,
and there can be many topologically distinct curve classes within a
single surface that could
each in principle support matter.

In fact, there is no clear {\it a priori} bound on the number of curves that
may support matter associated with any given kind of gauge group
factor  on a complex surface
of sufficiently complicated topology. In principle, a complex surface could have a very large number of
mutually disjoint complex curves, each appearing in the  anti-canonical class
and each supporting matter. We expect that there is a bound on
the set of surface types $S$ that can arise as divisors in complex
threefold
bases $B_3$, analogous to the bound on surfaces that can act as bases
for elliptically fibered Calabi-Yau threefolds. We do not, however, have an explicit
statement about the existence of such a bound.  Note that surfaces $S$  that cannot act as bases $B_2 = S$ for an
elliptically fibered Calabi-Yau threefold
can nonetheless arise as
divisors in a base $B_3$; for example, in the example above
of $B_3$ a $\P^1$ bundle over $\P^2$, with $n= 18$, one of the
divisors associated with a curve in the base $\P^2$ has the form of a
Hirzebruch surface $\F_{18}$, which cannot support an elliptically fibered Calabi-Yau
threefold.

In general, localized matter appears on a curve $C$ within a divisor
$D$ in $B_3$ when the Kodaira singularity type on $C$ exceeds that of
generic points on $D$.  For each gauge algebra type, for generic
coefficients in the Weierstrass model, this occurs when certain
coefficients $f_k, g_k$ vanish.  For example, on a divisor carrying a
type $III^{*}$ ($\ge_7$) codimension one singularity, matter will be
localized along the vanishing locus of $\hat{f}_3$, where the $(3, 5)$
$III^{*}$ singularity is enhanced to a $(4, 5)$ $II^{*}$ 
singularity.\footnote{The ``Kodaira type'' along the matter curve is
simply the type associated with the specified order of vanishing.
As observed in \cite{mt-singularities, Esole-Yau}, this does not
imply that the resolved Calabi--Yau manifold has a fiber of that
particular type in the codimension two locus.}
Like the 6D case of the $-7$ curve that carries an $\ge_7$ gauge
algebra and matter in the $\frac{1}{2} {\bf 56}$ representation, a
divisor $D \cong\P^2 \subset B_3$ with a normal bundle of $N=  -10H$
or $-11H$  carries an $\ge_7$ algebra; the matter locus is determined
by $\hat{f}_3 \in \Gamma ({\cal O} (12H + N))$, so the geometric matter lies on a
conic or pair of lines in the case $N= -10H$, and on a single line in
the class $H$ when $N= -11H$.

As an example of a situation where multiple curves support geometric
matter in a non-Higgsable cluster, consider the case where the divisor
$D = dP_3$ is a del Pezzo surface formed by
blowing up $\P^2$ at three points, giving three exceptional curves
$E_i$ with $E_i \cdot E_i = -1$, and three lines $L_{i j}= H - E_i -
E_j$.  The anti-canonical class is $- K = 3H - E_1 - E_2 - E_3
= L_{12}+ L_{13}+ L_{23}+ E_1 + E_2 + E_3$.
If we take the normal bundle of $D$ to be the line bundle
associated with the divisor $N=2E_1 -4L_{1 2}-4E_2 - 6L_{2 3}$, it is
straightforward to verify that the gauge algebra on $D$ is $\ge_7$,
and
\begin{equation}
 f_3 \in \Gamma ({\cal O} (-4K + N))
= \Gamma ({\cal O} (2H + 2E_1 + 2E_2 + 2E_3)) \,.
\end{equation}
It follows that matter is supported on the three disjoint curves
$E_i$.
This local geometry
can be embedded in a global threefold base $B_3$ that is a $\P^1$
bundle over $dP_3$, by simply taking $B_3$ to be the projectivization
of the line bundle $N$.  This example is one of the complete set of
possible $\P^1$ bundles over del Pezzo and generalized del Pezzo
surfaces  that were classified and studied in \cite{Anderson-WT}. 

Similar examples of non-Higgsable gauge groups associated with
geometric matter on one or more curves can be constructed for other
choices of gauge algebra.  Two examples of this type with a
non-Higgsable $\gsu_3$ realized through a type $IV$ singularity are
described
explicitly in \cite{ghst}.  In one of these examples, for instance,
$B_3$ is a $\P^1$ bundle over $dP_2$ with $N= E_1 + E_2 - L_{12}$,
which gives a type $IV$ singularity over the section $\Sigma_-$, which
is itself a $dP_2$ with normal bundle $N$.  We have
\begin{equation}
 g_2 \in \Gamma ({\cal O} (-6K + 4N))
= \Gamma ({\cal O} (2L_{12})) \,.
\end{equation}
Since $L_{12}= H - E_1 - E_2$ is a $-1$ curve, and hence a rigid
divisor within the surface $\Sigma_-$, this means that $g_2$  vanishes
to order 2 on $L_{1 2}$  and is a perfect square.  So the gauge
algebra is $\gsu_3$, and matter is localized on the curve $L_{12}$.

Other examples in which multiple curves within a divisor $D$
carry matter
can arise in situations where $D$ itself carries a gauge group but
also intersects with multiple other  divisors that also carry
nonabelian gauge group factors.  We  describe some explicit examples
of this kind in the following sections.

\section{Products of two factors} 

We now consider situations where a pair of divisors $D_a$ and $D_b$
intersect  on a curve $C_{a b}$ and
both carry non-Higgsable gauge groups.
Because the order of vanishing
of $f, g$ on $C_{a b}$ is at least the sum of
that on $D_a, D_b$, the same is true for $g$,
and the minimal orders of vanishing that give a
non-Higgsable gauge group factor are $(1, 2)$, we cannot have a product  containing anything larger than
a type $I_0^*$ $(2, 3)$
singularity.  
Furthermore, as noted above and in Appendix A, an $\gso_7$ algebra can
only arise when the orders of vanishing are $(2, 4)$, which would lead
to a $(4, 6)$ singularity when combined with an $\gsu_3$ component.
This restricts us to the eight possibilities
$\gsu_2 \oplus \gsu_2, \gsu_2 \oplus \gsu_3,
\gsu_3 \oplus \gsu_3, 
\gg_2 \oplus \gsu_2,  \gg_2 \oplus \gsu_3,
\gso_7 \oplus \gsu_2, 
\gso_8 \oplus \gsu_2,  \gso_8 \oplus \gsu_3$.

Some simple examples of product groups can be found by taking
bases of the form $B_3 = \P^1 \times B_2$ where $B_2$ contains a $(-2,
-3)$ non-Higgsable cluster; in these cases $B_3$ has a non-Higgsable
cluster associated with the lifts of the corresponding divisors, with
gauge algebra $\gsu_2 \oplus \gg_2$; note that the generic monodromy
condition is not modified in this case as the available set of
Weierstrass monomials simply increases in the product space.

Examples with a gauge algebra $\gsu_2 \oplus \gsu_3$,  corresponding
to the nonabelian part of the gauge group of the standard model of
particle physics,  were described in \cite{ghst}.

In \cite{Halverson-Taylor}, a systematic analysis of all models where
the base $B_3$ is a $\P^1$ bundle over one of the toric bases $B_2$
from \cite{toric} will be described.  These models contain a wide
range of examples of  non-Higgsable gauge groups, including
examples of five possible two-factor combinations:
\begin{equation}
\begin{array}{ccccc} 
\gsu_2 \oplus \gsu_2, &\hspace*{0.1in} & \gsu_2 \oplus \gsu_3,
&\hspace*{0.1in} & \gsu_3 \oplus \gsu_3, \\
\gg_2 \oplus \gsu_2, &\hspace*{0.1in} & 
\gso_7 \oplus \gsu_2
&\hspace*{0.1in} & 
\end{array}
\label{eq:pairings}
\end{equation}

In fact, these are the only possible algebras associated with
two-factor gauge products.  All other possibilities can be ruled out
by a local analysis.  

Considering the remaining possibilities in turn, first
assume that there exist divisors $A,
B$ that have nonzero intersection and that support gauge algebra
factors $\gsu_3 \oplus \gso_8$ or $\gsu_3 \oplus \gg_2$, with the
$\gsu_3$ supported on $A$.  We can choose local coordinates so that $z
= 0$ on  $A$ and $w = 0$ on $B$.
Then there must be a leading term in $g$ of the form $z^2 w^3$.  If
there were no such term than we would have a $(4, 6)$ singularity on
$A \cap B$.  If such a term exists, however, then $g/z^2|_{z = 0}$
cannot be a perfect square.  So $A$ cannot support a non-Higgsable
$\gsu_3$, giving a contradiction.  It follows that neither
$\gsu_3 \oplus \gso_8$ nor $\gsu_3 \oplus \gg_2$ can be realized in a
non-Higgsable cluster.

Now assume that $A, B$ intersect and
support non-Higgsable algebra factors $\gsu_2\oplus \gso_8$.  Using
the same coordinate system as above, the singularity on $A$ cannot be
type $IV$, since there would again be a leading term in $g$ of the
form $z^2 w^3$, and $g/w^3|_{w=0}$ would not be a perfect cube.  If
there is no term in $g$ of the form $z^2 w^3$ and we have a type $III$
singularity on $A$, then there must be a leading term in $f$ of the
form $zw^2$.  But then $f/w^2|_{w = 0}$ cannot be a perfect square, so
$B$ cannot support a non-Higgsable $\gso_8$.  Again we have a
contradiction, so there is no non-Higgsable gauge group containing a
connected pair of factors with algebra $\gsu_2\oplus \gso_8$.

Thus, all two-factor products are ruled out in non-Higgsable clusters
except for  those having the five algebras listed in
(\ref{eq:pairings}).

\section{More complicated ``quiver diagrams''}  

While in 6D, there are only two possible gauge algebras with multiple
components that can arise from a non-Higgsable cluster, in 4D the
structure is much richer. As in the case of matter, this occurs
because while all points on a curve represent the same element of
homology, on a surface there can be many distinct curves that are
mutually non-intersecting and are non-homologous.  A given divisor
that carries a non-Higgsable gauge group factor can thus intersect
with many other divisors, each of which carries a non-Higgsable gauge
group factor of its own, along a set of distinct curves.

This means that there is no obvious constraint that limits the number
of gauge factors that a given non-Higgsable gauge group factor can be
connected to through (geometric) matter in 4D models.  The set of
connected gauge group factors in the non-Higgsable cluster can thus
contain ``branchings'' where a single gauge group factor is connected
to three or more other factors.  Furthermore, since
a gauge factor can
in general be connected to two other factors, with no constraint other
than the limit on pairings from (\ref{eq:pairings}), it is  possible to have long chains connecting one branching point
to another, or to itself, which could in principle produce graphs of
arbitrary complexity in the absence of some global bound.

The graphs describing non-Higgsable clusters for 4D theories are
conveniently described by the standard  diagrammatic
convention of ``quivers,'' discussed in the physics literature in
\cite{Douglas-Moore}.  In a quiver diagram, each gauge group factor
is
represented by a node in a graph, and a directed arrow from a group
factor $G$ to a group factor $H$ corresponds to matter in a
bifundamental representation $(R_G,\bar{R}_H)$.  Since we are here
only focused on the geometric aspect of the gauge groups and matter
involved, all matter will be represented by bi-directional arrows,
corresponding to ${\cal N} = 2$ type matter in the 4D theory.

The local rules that we have derived here do not seem to place any
significant constraints on the complexity of quivers that can arise
from non-Higgsable geometries in F-theory constructions of 4D vacua.
It seems likely that, as for Calabi-Yau threefolds, 
the number of distinct topological classes of in
elliptically
fibered Calabi-Yau fourfolds is finite and hence that there are some
actual bounds on the complexity of possible quivers from compact
threefold bases $B_3$.  We leave a global analysis of these issues for
future work.  Here, we simply give a few examples in which the
branching and chain features just mentioned are realized explicitly.

\subsection{Branchings}
\label{sec:branchings}

\begin{figure}
\centering
\begin{picture}(200,110)(- 100,- 60)
\put(0,0){\makebox(0,0){$\gsu_2$}}
\put(50,50){\makebox(0,0){$\gg_2$}}
\put(-50,50){\makebox(0,0){$\gg_2$}}
\put(0,-50){\makebox(0,0){$\gsu_2$}}
\put(10,10){\vector( 1,1){30}}
\put( -10,10){\vector( -1,1){30}}
\put( 0,-10){\vector(0, -1){30}}
\put( 0,-40){\vector(0, 1){30}}
\put(40,40){\vector( -1, -1){ 30}}
\put( -40,40){\vector( 1, -1){ 30}}
\end{picture}
\caption[x]{\footnotesize The quiver diagram associated with a
  non-Higgsable cluster having gauge algebra $\gsu_2 \oplus \gsu_2
  \oplus \gg_2 \oplus \gg_2$, with bifundamental (geometrically
  non-chiral) matter connecting the first $\gsu_2$
component with the other three
gauge factors.}
\label{f:branched-diagram}
\end{figure}
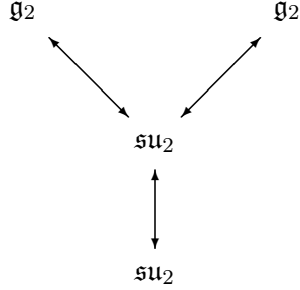

An explicit example of a base $B_3$ that gives a non-Higgsable cluster
exhibiting branching can be constructed as follows.  We begin with a
toric base $B_2$ chosen from the bases computed in \cite{toric},
characterized by a sequence of toric divisors (curves) $C_1, \ldots, C_9$
with self-intersections $- n_i$ 
\begin{equation}
 (- n_1, \ldots, - n_9) =
 (2, -3, -1, -3, -1, -4, -1, -4, 0)  \,.
\label{eq:example-branchings}
\end{equation}
We then construct $B_3$ as a $\P^1$ bundle over $B_2$ from the
projectivization of the line bundle $N= \sum_{i}a_i C_i$, where
\begin{equation}
 (- a_1, \ldots, - a_9) =
(0, 0,-1, 0, 1, 1, 3, 2, 5) \,.
\end{equation}
While the non-Higgsable cluster associated with this base can be
worked out in principle using the methods we have derived here, in
practice for a toric base like this it is generally easier to simply
analyze the orders of vanishing of $f, g$ on the various divisors
using the toric approach, which is easily automated.  To do this, the
divisors $C_i$ and the sections $\Sigma_\pm$ are represented as rays
$v_i$ in $N=\Z^{3}$.  The monomials in the Weierstrass model are then
the elements of the dual lattice $M = N^{*}$ that satisfy $\langle m,
v_i \rangle \geq -4, -6$ for $f, g$ respectively.  The orders of
vanishing on any divisor or curve can be determined by simply
considering the set of available monomials.  In this case, carrying
out this analysis shows that there is a type $III$ singularity on
$\Sigma_-$ carrying a $\gsu_2$ gauge algebra, and type $IV, I_0^*,$
and $I_0^*$ singularities on $C_4, C_6, C_8$ carrying gauge factors
$\gsu_2, \gg_2,$ and $\gg_2$.  The branched quiver diagram
representing this non-Higgsable cluster is depicted in
Figure~\ref{f:branched-diagram}. Note that for this particular
construction, there are no codimension three points where $f, g$
vanish to orders $4, 6$.  Many constructions exhibiting branching do
have such codimension three points.  Further examples of non-Higgsable
clusters exhibiting branching appear in the full set of $\P^1$ bundles
over $B_2$'s from \cite{toric}, and will be described further in
\cite{Halverson-Taylor}.

\subsection{Chains}

In six dimensions, there is only one situation (the $\gsu_2 \oplus
\gso_7 \oplus \gsu_2$ cluster) in which a non-Higgsable cluster contains a gauge
factor that intersects (carries jointly charged matter with) more than
one other gauge factor.  In four dimensions, however, this can happen
in a variety of ways, as discussed above.  In particular, there are
many local configurations in which a non-Higgsable gauge group can be
supported on a divisor $S$ that intersects two other divisors $S',
S''$ that both carry non-Higgsable gauge groups themselves.  This
opens the possibility of a long linear chain of connected gauge
factors in a non-Higgsable cluster.  Such chains cannot be ruled out
by the local analysis we have presented here.  Furthermore, some
exploration of the space of possible configurations shows that there
are global models in which relatively long non-Higgsable chains of
this kind can arise. 

As an explicit example, we demonstrate how it is possible in a simple
class of toric models to realize non-Higgsable clusters with gauge
algebras of the form $\gsu_2 \oplus \gsu_3 \oplus \gsu_3 \oplus \cdots
\gsu_3 \oplus \gsu_2$, where the number of $\gsu_3$ summands can go up
to at least 11.

Conceptually,  the idea of the examples we describe here is that we can
take a set of surfaces $S_i$ in $B_3$ to be a chain of $dP_3$ del
Pezzo surfaces,  each with normal bundle $N= - K$.  The surfaces will
be connected so that, for example,  $C_i=S_i \cap S_{i + 1}$ will be
the curve $E_1$ in $S_i$ and $L_{23}$ in $S_{i + 1}$, using the
notation of \S\ref{sec:single-matter}.  On each surface
in such a sequence we have $f_0, g_0 \in \Gamma ({\cal O} (0))$.  If we blow up
a curve on one such surface, it forces $f_0, g_0$ to vanish on that
surface.  This contributes to $\phi, \gamma$ on the adjoining
surfaces,  where $f_0, g_0$ must also vanish, etc.

To realize this explicitly in a 3D toric base $B_3$, consider the
following construction: we start with $\P^1 \times \F_m$, $m \leq 8$,
considered as a trivial $\P^1$ bundle over $\F_m$,
and denote by $\tilde{S}, F$ the divisors given by the lift of curves
in the base that are in the classes of the section with
self-intersection $+ m$ and the fiber.  We blow up on the curve
$\tilde{S}\cap F$, giving an exceptional divisor $E_1$; we repeat,
blowing up on the curve $E_1 \cap \tilde{S}$, then on the new curve $E_2
\cap \tilde{S}$ with $E_2$ the new exceptional divisor, etc., for a total of
$2m$ times.  We then blow up the curves $E_2 \cap \Sigma_-, \ldots,
E_{2m -2} \cap \Sigma_-$ in ascending order, and the curves $E_{2m -2} \cap
\Sigma_+, \ldots, E_2 \cap \Sigma_+$ in descending order, where
$\Sigma_\pm$ are two sections of the original
trivial $\P^1$ bundle, giving further exceptional divisors $E'_i,
E''_i$ for $i \in \{2, \ldots, 2m -2\}$.  This geometry contains
within it a linear chain of $2m -3$ del Pezzo surfaces $dP_3$ with
normal bundles $N= - K$ as described above; these are the proper
transforms of the surfaces $E_2, \ldots, E_{2m -2}$.  We then blow up
the curve $E_2 \cap E'_2$.  This leads to $(4, 6)$ singularities over
$E_i \cap E'_i, E_i \cap  E'_{i -1}$ for $i = 3, \ldots, 2m -4$, which are
resolved by blowing up these curves.  The final geometry has no $(4,
6)$ divisors or curves when $m \leq 8$,
and can be analyzed most easily by toric
methods to have a non-Higgsable
gauge  algebra $\gsu_2 \oplus \gsu_3 \oplus \cdots
\oplus \gsu_3 \oplus \gsu_2$  on the chain of divisors $E_2 \ldots,
E_{2m -2}$.
In the case $m = 8$, we thus have a chain of 13 non-Higgsable gauge
group factors, including 11 copies of $\gsu_3$.
For $8 < m <12$, the situation becomes more complicated as an $\ge_8$
singularity with  $(4, 6)$ matter curves develops on the surfaces $S,
\tilde{S}$; we do not analyze the details of these geometries here.

\begin{figure}
\centering
\begin{picture}(200,210)(- 100,- 110)
\put(-120,0){\makebox(0,0){$\gsu_2 \longleftrightarrow
\gsu_3 \longleftrightarrow
\gsu_3 \longleftrightarrow
\gsu_3 \longleftrightarrow \gsu_2
$}}
\put(-120, -25){\makebox(0,0){(A)}}
\put(120,15){\makebox(0,0){ \includegraphics[width=7cm]{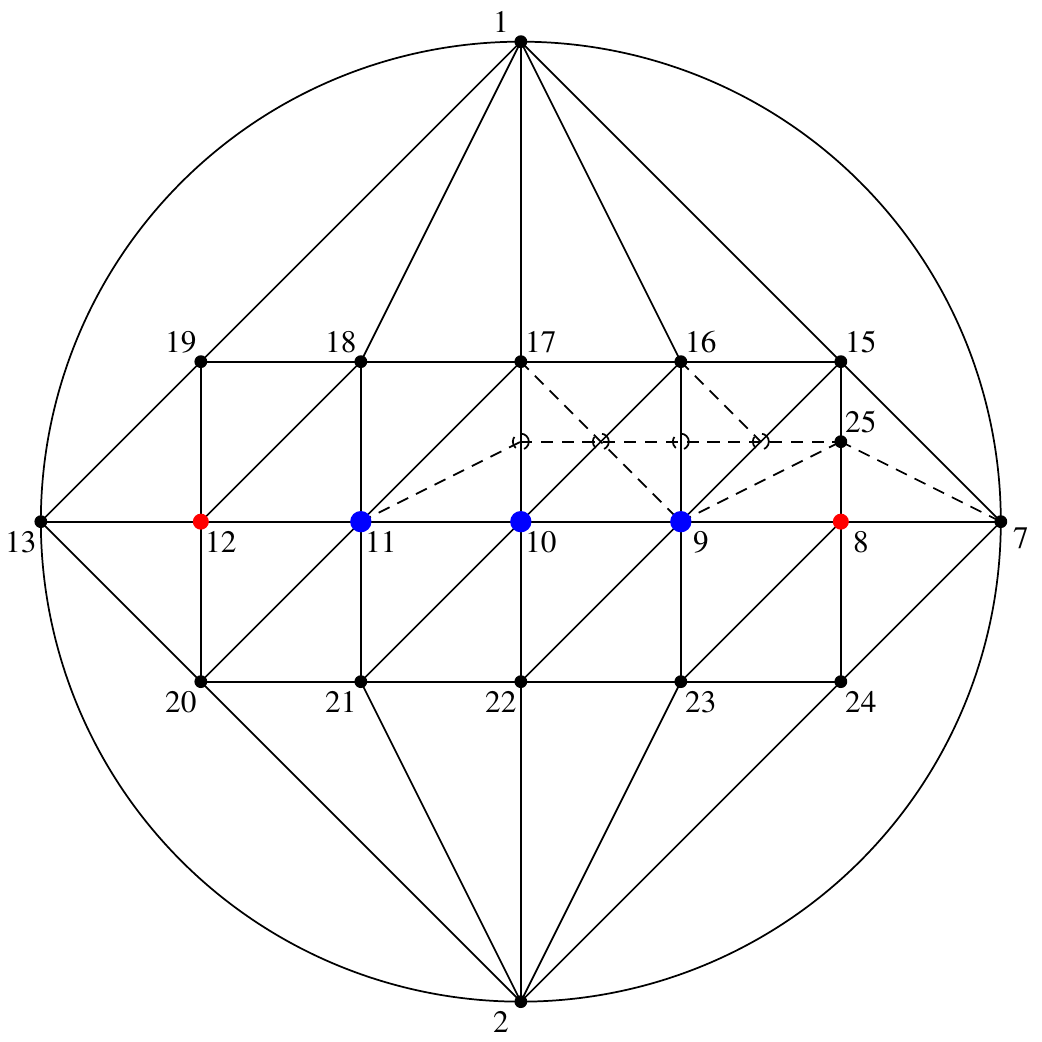}}}
\put(120, -95){\makebox(0,0){(B)}}
\end{picture}
\caption[x]{\footnotesize A global model with a non-Higgsable geometry
  giving rise to a linear chain in the quiver structure of the gauge
  algebra, $\gsu_2 \oplus \gsu_3 \oplus \gsu_3 \oplus \gsu_3 \oplus
  \gsu_2$.  (A) the quiver diagram of the chain, (B) A schematic
  depiction of the triangulation of the 3D toric fan describing the
  global geometry, formed from a sequence of blowups at points $7 (E_1),
  \ldots,$ from an initial fan describing the threefold $\P^1 \times
  \F_4$. 
(Note that points associated with the toric rays $v_3$-$v_6$
and $v_{14}$ are not shown.)
 Large blue dots represent divisors supporting an $\gsu_3$
  gauge summand, smaller red dots are divisors supporting an $\gsu_2$
  gauge algebra.  Open circles represent $(4, 6)$ curves that must be
  blown up to divisors after the blow-ups up to $v_{25}$.
Note that before blowing up to $v_{25}$, the divisors associated with
points $9, 10, 11$ are connected del Pezzo $dP_3$ surfaces
(as can be seen from the structure of solid lines).
Similar constructions are possible with up to (at least) 11
  factors of $\gsu_3$ in the linear chain.  }
\label{f:chain}
\end{figure}
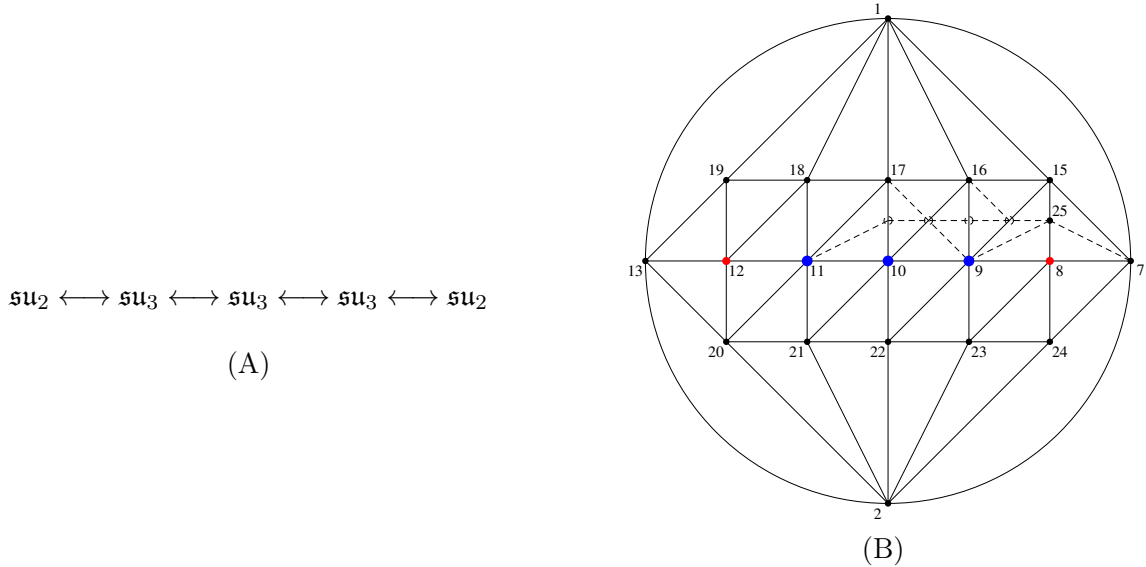

This construction is most easily visualized in the toric language,
where the toric divisors can be defined through the fan
\begin{eqnarray}
(\Sigma_\pm)\hspace*{0.1in}  v_{1, 2} & = & (0, 0, \pm 1)\\ 
(\tilde{S})\hspace*{0.1in} v_3 & = & (0, 1, 0) \\
\hspace*{0.1in} v_4 & = & (1, 0, 0) \\
({S})\hspace*{0.1in} v_5 & = & (0,  -1, 0) \\
(F)\hspace*{0.1in} v_6 & = & (  -1, - m, 0) \\
(E_i)\hspace*{0.1in}v_{6 + i} & = & (-1, - m + i, 0), \; \;
1 \leq i \leq 2m \\
(E'_i)\hspace*{0.1in}v_{5 + 2m + i} & = & (-1, - m + i, 1), \; \;
2 \leq i \leq 2m -2 \\
(E''_i)\hspace*{0.1in}v_{2 + 6m - i} & = & (-1, -
m + i, -1), \; \;2 \leq i \leq 2m -2 \\ 
\hspace*{0.1in}  v_{6m -1 +i} & = & (-2, -2m + 2 i , 1), \; \;
2 \leq i \leq 2m-4\\
\hspace*{0.1in}  v_{8m-6 +i} & = & (-2, -2m + 2 i + 1, 1), \; \;
2 \leq i \leq 2m-5
\end{eqnarray}
The triangulation associated with the cone structure for this fan  is
shown schematically in Figure~\ref{f:chain}
for the case $m = 4$.

This family of examples illustrates the possibility that long chains
of connected gauge group factors can arise in a 4D non-Higgsable
cluster geometry.

\subsection{Loops}

Another complication that can arise in a 4D non-Higgsable cluster is
the presence of a closed loop in the quiver diagram.  Locally, such a
loop looks much like  the chains described in the previous subsection,
and there is no way to rule out such loops based on purely local
considerations.

\begin{figure}
\centering
\begin{picture}(200,200)(- 100,- 100)
\put(-125,25){\makebox(0,0){$\gsu_2$}}
\put(-125,-25){\makebox(0,0){$\gsu_2$}}
\put(-75,25){\makebox(0,0){$\gsu_2$}}
\put(-75,-25){\makebox(0,0){$\gsu_2$}}
\put(-75,-15){\vector( 0,1){30}}
\put(-125,-15){\vector( 0,1){30}}
\put(-75,15){\vector( 0,-1){30}}
\put(-125,15){\vector( 0,-1){30}}
\put(-115,25){\vector(1, 0){30}}
\put(-115,-25){\vector(1, 0){30}}
\put(-85,25){\vector(-1, 0){30}}
\put(-85,-25){\vector(-1, 0){30}}
\put(-100, -50){\makebox(0,0){(A)}}
\put(100, 15){\makebox(0,0){ \includegraphics[width=7cm]{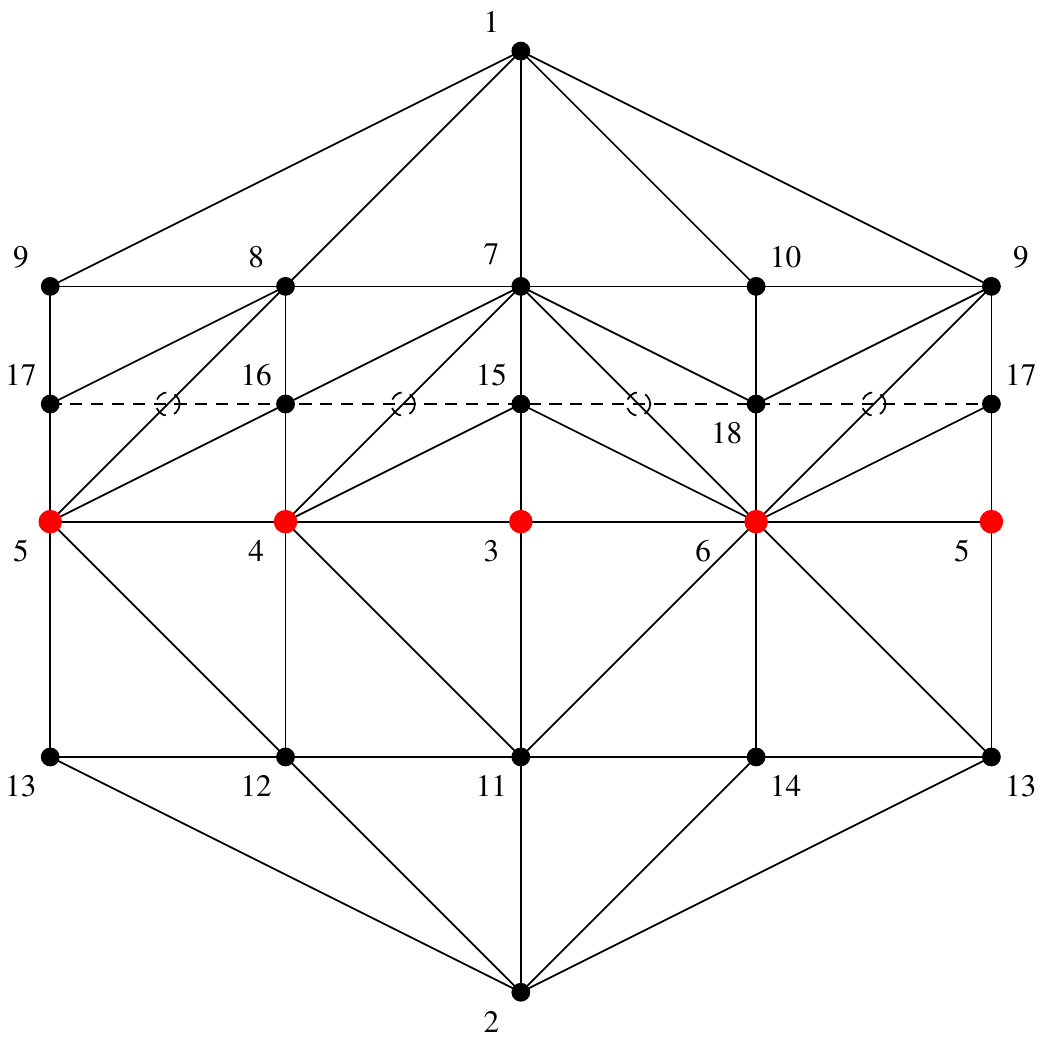}}}
\put(100, -100){\makebox(0,0){(B)}}
\end{picture}
\caption[x]{\footnotesize A global model with a non-Higgsable geometry
giving rise to a loop in the quiver structure of the gauge algebra.
(A) the quiver diagram of the loop, (B) A schematic depiction of the
triangulation of the 3D toric fan describing the global geometry,
formed from a sequence of blowups at points $7, \ldots,$ from an
initial fan describing the threefold $\P^1 \times \P^1 \times\P^1$.
(Note that the points on the left should be identified with the points on the right with the same labels.)
}
\label{f:loop}
\end{figure}
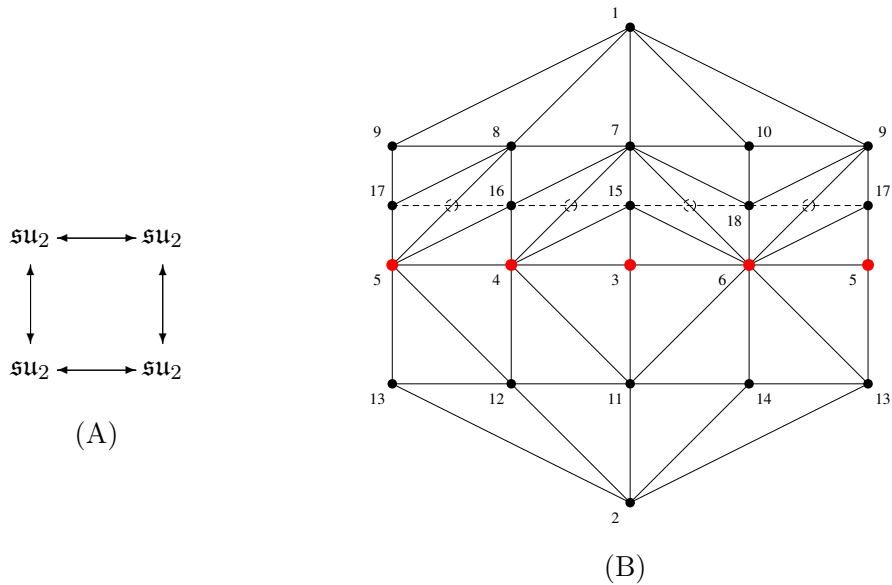

We have identified a number of examples where a loop arises in
a non-Higgsable cluster geometry.  In one simple example, there is a loop
consisting
of four $\gsu_2$ factors, so that the total gauge algebra is $\gsu_2
\oplus \gsu_2 \oplus \gsu_2 \oplus \gsu_2$, with matter curves
supporting matter in the bifundamental of each adjacent pair of gauge
group factors, as well as between the initial and final factors.  This
example can be constructed as follows: beginning with $\P^1 \times
\P^1 \times \P^1$, with divisors $X_\pm, Y_\pm, Z_\pm$ associated with
two distinct points on each of the three $\P^1$'s, and labeling $D_i =
Y_+, X_+, Y_-, X_-$ for $i = 1, 2, 3, 4$, we first blow up the curves
$Z_\pm \cap D_i$, giving new exceptional divisors $E_{i \pm}$.  We
then blow up on the curves $E_{i  +}\cap D_i$.
Blowing up four more $(4, 6)$
curves gives us a good F-theory base with no $(4, 6)$ divisors,
curves, or points.  In the final geometry, the proper transforms of
the initial divisors $D_i$ each carry an $\gsu_2$ gauge factor, and
they are connected in a cyclic chain as described above.  This can all
be done in the toric language, starting with the fan spanned by the
rays $v_1 = (0, 0, + 1), v_2 = (0, 0, -1), v_3 = (0, 1, 0), v_4 = (1,
0, 0), v_5 =  (0, -1, 0), v_6 = (-1, 0, 0)$, corresponding to $Z_+,
Z_-, Y_+, X_+, Y_-, X_-$, and then blowing up on the appropriate edges
of the toric fan in the sequence described above.
A schematic picture of the triangulation of the final fan, with
vertices labeled in the order of blowups, is given in
Figure~\ref{f:loop}.

\section{Conclusions}
\label{sec:conclusions}

\subsection{Summary and open questions} 

We have initiated a systematic analysis of geometric non-Higgsable
clusters that can arise in threefolds $B_3$ for F-theory
compactifications that give ${\cal N} = 1$ supergravity theories in
four dimensions.  These structures describe gauge groups and matter
that cannot be Higgsed at the geometric level by deformation of
complex structure moduli, and which therefore arise at generic points
in the moduli space of the corresponding Calabi-Yau fourfold.  More
work must be done to ascertain whether these geometrically
non-Higgsable structures are truly present in the low-energy
supergravity theory of any specific F-theory vacuum.  The presence of
G-flux, the corresponding superpotential, and extra degrees of freedom
not yet incorporated systematically into F-theory may affect this
conclusion.  Unless one of these factors causes a generic change in
qualitative behavior, however, it seems that the non-Higgsable
structures we have analyzed here will be generic features of F-theory
vacua.

We have developed a set of local equations that govern the possible
non-Higgsable structures  that can appear on a combination of divisors
in the F-theory base manifold.  For toric bases, a straightforward
monomial analysis can be used to determine  non-Higgsable
structures in any specific case.  

The set of individual gauge factors that can arise in a non-Higgsable
cluster is quite limited, and contains only nine distinct possible
simple Lie algebras.  Similarly, the number of products of two gauge
factors that can arise is also quite limited, and consists of only
five possibilities, including that of the nonabelian part of the
standard model $SU(3) \times SU(2)$, which was analyzed specifically
in \cite{ghst}.  Unlike in 6D, however, the topological structure of
the gauge groups in a non-Higgsable cluster, which can be depicted in
a quiver diagram, can apparently be quite complicated.  From the local
analysis there is no constraint that prohibits branching, loops, or
long linear chains of gauge factors.  Indeed, we have identified
explicit global geometries that contain each of these three features.
An interesting open question is the extent to which global constraints
limit the complexity of the graph structures that can arise for large
non-Higgsable clusters.  Indeed, since unlike for elliptically fibered
Calabi-Yau threefolds, there is as yet
no proof that the number of distinct
topological types of elliptically fibered Calabi-Yau fourfolds is
finite, there is no clear argument at this time that would bound the
size of possible non-Higgsable clusters for 4D F-theory
compactifications.

One issue that we have not addressed here is whether geometries with
codimension three
$(4, 6)$ singularities at points need to be treated in any special
way.  Such singularities cannot be blown up without tuning additional
complex structure moduli.  These therefore do not represent
additional branches in the space of fourfolds, unlike
$(4, 6)$ singularities in codimension two.  Like codimension two $(4, 6)$ singularities,
however, which correspond to superconformal field theories  coupled to
gravity,
the codimension three singularities may also represent some kind of
more exotic gravitationally coupled theory.

\subsection{Classifying Calabi-Yau fourfolds}

The classification of non-Higgsable clusters on 2D bases
\cite{clusters} has enabled a systematic analysis of the space of
elliptic Calabi-Yau threefolds, giving a fairly clear global picture
of the space of 6D F-theory models \cite{toric, Hodge,
Martini-WT,
  Johnson-WT, Wang-WT}.  
It is hoped that the results presented here will
similarly provide a useful tool for exploring the space of 3D bases
for F-theory compactifications to four dimensions.  Such analysis of
elliptic Calabi-Yau fourfolds, however, will be significantly more
complicated than the corresponding analysis for elliptic Calabi-Yau
threefolds.  For elliptic Calabi-Yau threefolds, the minimal model program
\cite{bhpv} and the work of Grassi \cite{Grassi} give a simple characterization
of the set of possible bases as blowups of the Hirzebruch surfaces
$\F_m, m \leq 12,$ $\P^2$, and the Enriques surface.  For elliptic Calabi-Yau fourfolds, where the
base is a complex threefold, no such simple characterization of
minimal bases is known; constructing such a classification using the
minimal model approach (Mori theory) in higher dimensions is an
interesting open problem.  The analysis of non-Higgsable clusters
given here may provide some guidance in attempting to systematically
address this problem.

One physical motivation for attempting a systematic global
characterization of the space of possible
elliptic Calabi-Yau fourfolds and
associated non-Higgsable clusters is the goal of identifying generic
features or specific constraints that F-theory places on 4D
supergravity theories.  In six dimensions, F-theory geometry places
various constraints on the possible effective theories that can be
realized.  Some of these constraints are understood in the physical
theory as anomaly constraints, while other constraints place
additional consistency conditions on the low-energy theory
\cite{KMT-II, Seiberg-WT}.  F-theory geometry also places strong
constraints on the compactification geometry that have manifestations
in the low-energy 4D supergravity theory.  Some initial exploration of
such constraints was carried out in \cite{Grimm-WT, Anderson-WT}, but
it seems likely that as our understanding of the global space of
F-theory vacua matures further insights into the constraints produced
on low-energy theories will emerge.

The approach of analyzing elliptic fibrations through the geometry of
the base provides a complementary approach to the long-studied toric
approach to describing Calabi-Yau manifolds as hypersurfaces in toric
varieties pioneered by Batyrev \cite{Batyrev} and the complete
intersection (CICY) approach taken in \cite{CICY}.  While both these
approaches give a large class of Calabi-Yau manifolds, and can be used
to study elliptic Calabi-Yau's  (see {\it e.g.}
\cite{Candelas-font, Candelas-cs,
Braun-gk1, Klevers-mopr, Braun-gk2, Gray-hl, Gray-hl-2}, the classification of bases
using non-Higgsable clusters is in principle both a simpler approach
as it reduces the complexity of the geometry involved, and a more
complete approach as it is in principle possible to describe all
elliptic Calabi-Yau's from this point of view.  We expect that the
complementary insights provided by these different approaches, all of
which are currently under active investigation, will provide many new
insights into the geometry of elliptic Calabi-Yau fourfolds in the
near future.

\subsection{Physical consequences of non-Higgsable clusters} 
\label{sec:applications}

At the present time, F-theory represents one of the most general
approaches to constructing vacuum solutions of string theory for which
analytic tools are available.  While, unlike in 6D, F-theory in its
current formulation does not seem to in any sense cover the full space
of 4D string vacua, the vacua formed from F-theory seem to represent a
much larger and broader sample than those 4D vacua constructed from
many other approaches.  For example, F-theory vacua with a smooth
heterotic dual are a small subset of the full set of possible
F-theory constructions \cite{Anderson-WT}.  And since the Hodge
numbers of generic elliptic Calabi-Yau fourfolds are much larger than
those of threefolds, the number of flux vacua formed from F-theory
constructions is estimated to be much much larger than those from
other constructions such as IIB flux vacua \cite{Denef-F-theory}.
On the other hand, the apparently infinite number of topologically
distinct type IIA flux
vacua
\cite{IIA} and the potentially even greater number of non-geometric
string compactifications may provide even larger families of vacua
than those realized through standard F-theory compactifications.
F-theory also, however, provides a window on the nonperturbative dynamics of
string vacua in a way that is not available from other string
constructions that require a weak coupling limit.  Given these
observations, it seems that F-theory gives one of the best pictures we
have so far for the behavior of a ``generic'' class of nonperturbative
string vacua.  Assuming that issues such as G-flux do not
substantially change the structure of the gauge factors that are
forced by non-Higgsable clusters in the F-theory geometry, a
suggestive picture emerges of the physics of a ``typical'' F-theory
vacuum in the landscape.  In particular, as emphasized also in
\cite{ghst}, non-Higgsable clusters provide a mechanism that may make
light matter fields and gauge symmetries a natural consequence of
generic string compactifications, without requiring any special
tuning.

Non-Higgsable structures seem to be highly prevalent in F-theory
vacua.  In six dimensions, of the more than 100,000 possible base
manifolds studied in \cite{toric, Martini-WT} that support elliptic
Calabi-Yau threefolds, only 27 lack non-Higgsable gauge groups.
Typical elliptic threefolds with large Hodge numbers have many
non-Higgsable clusters, with numerous factors of the gauge algebras
$\ge_8, \gf_4, \gg_2 \oplus \gsu_2$.  We expect a similar story to
hold for fourfolds, though the types of factors that are typical in 4D
models has not yet been systematically analyzed.  In particular, we
expect that for fourfolds with large Hodge numbers, which are likely
to give rise to the largest number of distinct flux vacua, there will
typically be many non-Higgsable gauge group factors.  Some initial
systematic investigation in this direction will be presented in
\cite{Halverson-Taylor}.

While the possible structures that can arise in non-Higgsable clusters
for 4D models may be quite complicated, the set of possible gauge
groups, and in particular the products of two gauge groups that can
appear in these clusters, is actually quite limited.  Specifically,
the nonabelian part of the standard model gauge group, $SU(3) \times
SU(2)$, is one of only five possible two-summand Lie algebra
structures that can arise from non-Higgsable clusters.  If we assume
that matter, and at least two nonabelian gauge factors, are the
minimal necessary components for ``interesting'' ({\it i.e.},
anthropic) physics in the landscape, then the nonabelian part of the
standard model arises as simply one of five natural minimal
possibilities that may arise throughout the landscape.  While this is
certainly suggestive, and may provide an alternative framework for
F-theory phenomenology to the well-studied F-theory GUT approach
\cite{Donagi-Wijnholt, bhv, bhv2, Heckman-review, Weigand-review}, many
significant questions must be answered to provide realistic models of
particle physics from this approach.  The abelian $U(1)$ factor in the
standard model, for example, must either be tuned by hand, or must
also arise in a non-Higgsable fashion.  While the latter possibility
has been shown to be possible in 6D models \cite{Martini-WT,
  Morrison-Park-Taylor}, such a mechanism is not yet well understood
in four dimensions.  It is also necessary to understand better how
geometric non-Higgsability relates to the field theory description of
the low-energy theory, and for a more realistic model further
structure such as the Yukawa couplings would need to be computed in
any specific geometry with the proper gauge groups and matter content.
Note that while of course the $SU(2)$ of the standard model seen in
nature is broken by the Higgs field, this could in principle occur
even in a 4D F-theory model with the $SU(2)$ in a non-Higgsable
cluster, if there is an appropriate geometrically non-Higgsable matter
field charged under the $SU(2)$ that acquires a negative mass through
radiative corrections after SUSY is broken; further discussion and
analysis of the non-Higgsable $\gsu_3\oplus \gsu_2$ structure
appears in \cite{ghst}.

If we assume that all nonabelian gauge groups and matter arising in nature come
from generic ({\it i.e.} non-Higgsable) structures, then the structure
of non-Higgsable gauge group factors  would also place interesting constraints
on dark matter.  One possibility for dark matter is a non-Higgsable
cluster (or multiple clusters) with one or more gauge group factors
that are completely disconnected from the standard model; such
disconnected dark matter sectors have also been considered in the
F-theory GUT literature \cite{bhv2}.  In the context of non-Higgsable
clusters, this would
correspond to hidden sector dark matter 
with
specific possible gauge groups and matter content.  In the simplest
cases, this could be simply an additional supersymmetric $\gsu_2,
\gsu_3, \gg_2, \gso_7,
\gso_8, \gf_4, \ge_6, \ge_7$
or $\ge_8$ sector with a spectrum of glueballs that would interact
only gravitationally with ordinary matter.  Another possibility
for dark matter can arise if the nonabelian standard model components $SU(3) \times
SU(2)$ lie in a non-Higgsable cluster, but with other gauge factors
connected in the quiver diagram.  For example, the standard model
factors could arise as part of a non-Higgsable cluster with gauge
algebra $SU(3) \times SU(2) \times G$, with additional matter charged
under the $SU(2)$ and $G$ factors; this would correspond to a weakly
interacting dark matter sector.  An interesting consequence
is that if such a
dark matter sector arises from a non-Higgsable cluster then it
would have to be associated with an internal gauge group $G$ that
would be restricted to have one of the gauge algebras $\gsu_2, \gsu_3,
\gg_2,$ or $\gso_7$.  In fact, these are the only
possibilities for gauge algebras that can connect to the $\gsu_2$ of the
standard model in any non-Higgsable cluster.  The discovery of
matter charged under a hidden gauge group
in this family would thus fit
naturally with the predictions of a generic F-theory model.  On the
other hand, discovery of weakly interacting matter with, for example,
an additional
$SU(N)$ gauge group sector where $N> 3$ would rule out the hypothesis that
  the low-energy spectrum seen in nature arises from non-Higgsable
  geometric structures in generic F-theory models.

Combining these features, the simplest picture of the ``typical''
F-theory vacuum that we have available at this time would be a model
where the visible light fields consist of one or more single or
multi-factor gauge groups with or without matter, with only five
possibilities including $ SU(3) \times SU(2)$ for two-factor visible
gauge group products carrying jointly charged matter.  There could also be dark matter sectors corresponding to
other additional, completely decoupled, non-Higgsable clusters, or
additional sectors that come from the same cluster and that could be
charged for example under an $SU(2)$
in an $SU(3) \times SU(2)$ product as well as another hidden gauge
group that would have to have the algebra $\gsu_2, \gsu_3, \gg_2$, or
$\gso_7$ as just discussed above.  Clearly, supersymmetry breaking,
which we have completely ignored here, would need to be incorporated
in any realistic model.  This would also give rise to possible
pseudoscalar axion-like fields from the lifting of scalar moduli from
the supersymmetric model.  The structure of light fields that we
see in the observed universe is not too different from this highly
simplified picture of what we might expect from a generic F-theory
vacuum.  At our current state of understanding, this picture of the
typical F-theory vacuum is still quite incomplete and cannot yet be
used to make specific predictions for low-energy non-supersymmetric
physics.  While the analysis of geometric non-Higgsable clusters in
this paper is based on rigorous mathematical reasoning, the connection
between the underlying geometry and low-energy physics is not as
direct in 4D F-theory models as in six dimensions, where the
low-energy physics precisely mirrors the geometry.  Much more work
must be done to understand the role of G-flux and 7-brane world-volume
degrees of freedom in
F-theory, and to incorporate supersymmetry breaking into
supersymmetric 4D F-theory vacuum models.  It seems possible, however,
that even as our understanding improves the non-Higgsable clusters
described here may continue to play an important and perhaps
predictive role in describing the generic properties of 4D
supersymmetric vacua of F-theory.

\vspace*{0.1in}

{\bf Acknowledgements}: We would like to thank Lara Anderson,
Mboyo Esole,
Antonella Grassi,
Jim Halverson, Jonathan Heckman,
Sam
Johnson, Shamit Kachru, Shu-Heng Shao, Eva Silverstein, Tracy Slatyer, and Yinan Wang for helpful discussions and the referee for helpful comments.  
We would also
like to thank the Aspen Center for Physics, where the initial stages
of this work were carried out.  The research of W.T.\ is supported by
the U.S.\ Department of Energy under grant Contract Number
DE-SC00012567.  The research of D.R.M.\ is supported by by the
National Science Foundation under grant PHY-1307513.  
We thank the Aspen
Center for Physics for hospitality and partial support by the National
Science Foundation Grant No. PHYS-1066293.  

\appendix

\section{The gauge algebra of maximally Higgsed models}

In this appendix we discuss the monodromy conditions for F-theory seven-branes
of type $I_0^*$.  If $D=\{z=0\}$ describes the location of the brane (in local
coordinates),
then $f$, $g$, and $\Delta$ have orders  ${}\geq2$, ${}\geq3$, and $6$ along $D$,
respectively.  We let $\hat{f}=(f/z^2)|_{\{z=0\}}$, 
$\hat{g}=(g/z^3)|_{\{z=0\}}$ and 
$\hat{\Delta}=(\Delta/z^6)|_{\{z=0\}}$ (all in the local coordinate chart).
These are sections of line bundles 
$$\mathcal{O}_D(-2kK_B-kN_{D/B})= \mathcal{O}_D(-2kK_D+kN_{D/B})$$
for $k=2,3,6$, respectively.

The monodromy and gauge algebra are now determined by the behavior of the
cubic polyomial
\[ x^3+\hat{f}x+\hat{g}.\]
\begin{itemize}
\item If there are sections $\alpha,\beta\in \Gamma(\mathcal{O}_D(-2K_D+N_{D/B}))$ such that
$\hat{f}=-(\alpha^2+\alpha\beta+\beta^2)$ and
$\hat{g}=\alpha\beta(\alpha+\beta)$,
then
\[ x^3+\hat{f}x+\hat{g}=(x-\alpha)(x-\beta)(x+\alpha+\beta)\]
and the gauge algebra is $\mathfrak{so}(8)$.  To ensure that the seven-brane
has type $I_0^*$, we also need
\[ \hat{\Delta}=-(\alpha-\beta)^2(2\alpha+\beta)^2(\alpha+2\beta)^2\]
to be not identically zero.  That is, three things must be avoided:
\begin{itemize}
\item[(i)] $\beta=\alpha$, which would imply $\hat{f}=-3\alpha^2$
and $\hat{g}=2\alpha^3$.
\item[(ii)] $\beta=-2\alpha$, which would imply $\hat{f}=-3\alpha^2$
and $\hat{g}=2\alpha^3$.
\item[(iii)] $\beta=-\frac12\alpha$, which would imply
$\hat{f}=-\frac34\alpha^2$, $\hat{g}=-\frac14\alpha^3$.
\end{itemize}
\item If there are sections $\lambda\in \Gamma(\mathcal{O}_D(-2K_D+N_{D/B}
))$ 
and $\mu \in \Gamma(\mathcal{O}_D(-4K_D+2N_{D/B}
))$ such that 
$\lambda^2-4\mu$ is not the square of a section of $\mathcal{O}_D(-2K_D+N_{D/B})$ and such that
$\hat{f}=\mu-\lambda^2$ and $\hat{g}=-\lambda\mu$, then
\[x^3+\hat{f}x+\hat{g}=(x-\lambda)(x^2+\lambda x+\mu)\]
and the gauge algebra in $\mathfrak{so}(7)$.  To ensure that the seven-brane
has type $I_0^*$, we also need
\[ \hat{\Delta}=(\mu+2\lambda^2)^2(4\mu-\lambda^2)\]
to not vanish identically.
\item In all other cases, the gauge algebra is $\mathfrak{g}_2$.
\end{itemize}

We point out two particular
ways to solve these constraints (although these solutions are not
the most general ones possible).  
\begin{itemize}
\item[]\hspace*{-0.38in}{\bf Solution 1.}\\
If the spaces of sections  
$\Gamma(\mathcal{O}_D(-2K_D+N_{D/B}))$,
$\Gamma(\mathcal{O}_D(-4K_D+2N_{D/B}))$ and
$\Gamma(\mathcal{O}_D(-6K_D+3N_{D/B}))$,
are all one-dimensional and $\hat{u}\in \Gamma(\mathcal{O}_D(-2K_D+N_{D/B}))$
is not identically zero, then there are constants $A$ and $B$ so that
 $\hat{f}=A\hat{u}^2$ and $\hat{g}=B\hat{u}^3$.  There are then
constants $r_1$, $r_2$, $r_3$ such that the constant polynomial
$X^3+AX+B$ can be factored into linear factors:
\begin{equation}
\label{eq:constant factors}
X^3+AX+b=\prod_{i=1}^3 (X-r_i)
\end{equation}
and there is a corresponding factorization of a section of
$\mathcal{O}_D(-6K_D+3N_{D/B})$, namely,
\begin{equation}
\label{eq: section factors}
x^3+\hat{f}x+\hat{g} = \prod_{i=1}^3 (x-r_iu).
\end{equation}
In this case, we
get gauge algebra $\mathfrak{so}(8)$ for any choice of $\hat{f}$ and
$\hat{g}$ provided that $\hat{\Delta}\not\equiv0$.
\item[]\hspace*{-0.38in}{\bf Solution 2.}\\
If $\hat{g}\equiv0$ and $\hat{f}$ is not a square,
then the gauge algebra must be $\mathfrak{so}(7)$.  For in this case,
we can set $\lambda=0$, $\mu=\hat{f}$ to satisfy the criterion given above.
Note that $\hat{\Delta}=4\mu^3\not\equiv0$ since $\mu$ is not a square.
\end{itemize}

We now consider what monodromies can occur in the case of an F-theory
seven-brane of type $I_0^*$ in a model which has been maximally Higgsed.  Fixing the
base $B$ and the divisor $D$ at which the seven-brane is located,
there are restriction maps
\[ \rho_k:\Gamma(\mathcal{O}_B(-2kK_B+kD))\to\Gamma(\mathcal{O}_D(-2kK_D+kN_D))\]
(identifying $\mathcal{O}_D(D)$ with $\mathcal{O}_D(N_D)$).
If $f$ and $g$ are generic,  {\it i.e.}, the model is maximally Higgsed,
then  $\hat{f}$ will be a generic element of
the image of $\rho_2$, while $\hat{g}$ will be a generic element of the 
image of $\rho_3$.  Here is how we will use the ``maximally Higgsed''
property:
if we scale $(\hat{f},\hat{g})\to(c_1\hat{f},c_2\hat{g})$, we should
obtain the same gauge algebra for general constants $c_1$, $c_2$.
(If not, then further Higgsing is possible by scaling $f$ and $g$.)

Consider first the case
that the gauge algebra is $\mathfrak{so}(8)$, and let
 $\alpha,\beta\in \Gamma(\mathcal{O}_D(-2K_D+N_{D/B}))$ be  sections
such that $\hat{f}=-(\alpha^2+\alpha\beta+\beta^2)$ and 
$\hat{g}=\alpha\beta(\alpha-\beta)$. Suppose that $\alpha$
and $\beta$ are linearly independent in the complex vector space
$\Gamma(\mathcal{O}_D(-2K_D+N_{D/B}))$.  We first remark that
this implies that $\{\alpha^2,\alpha\beta,\beta^2\}$ are
linearly independent in 
$\Gamma(\mathcal{O}_D(-4K_D+2N_{D/B}))$ and that 
$\{\alpha^2\beta,\alpha\beta^2\}$ are linearly independent in
$\Gamma(\mathcal{O}_D(-6K_D+3N_{D/B}))$.  This is because a linear dependence
relation among powers of degree $N$,
 $\sum K_j \alpha^j \beta^{N-j}\equiv0$ with constant
coefficients $K_j$, would lead to a linear
dependence relation among powers of degree $1$
by factoring the homogeneous polynomial 
$\sum K_j \xi^j \eta^{N-j}\in \mathbb{C}[\xi,\eta]$ into homogeneous linear
factors and choosing a factor which vanishes upon subtituting $\alpha$ for
$\xi$ and $\beta$ for $\eta$.

Since $\alpha$ and $\beta$ are assumed to be
linearly independent, none of
 $\alpha$,
$\beta$, and $\alpha+\beta$ can vanish identically.  Thus, the locus 
$\{\hat{g}=0\}$ must decompose as a union of $\{\alpha=0\}$,
$\{\beta=0\}$, and $\{\alpha+\beta=0\}$.  After generic scaling
$(\hat{f},\hat{g})\to(c_1\hat{f},c_2\hat{g})$ we get the same
locus:
\[ \{c_2\hat{g}=0\}=\{\hat{g}=0\}.\]
Thus, in order to get the same kind of decomposition,
after permuting $\{\alpha,\beta,-\alpha-\beta\}$ if necessary
(which can be achieved
with a linear transformation on the span of $\alpha$ and $\beta$),
we can assume that the sections $\alpha'$ and $\beta'$ that are needed
after scaling take the form
\[ \alpha'=c_3\alpha; \quad \beta'=c_4\beta.\]
Now we get several equations from the linear independence of
$\{\alpha^2,\alpha\beta,\beta^2\}$ and $\{\alpha^2\beta,\alpha\beta^2\}$.
Since $c_1(\alpha^2+\alpha\beta+\beta^2)=c_3^2\alpha^2+c_3c_4\alpha\beta
+c_4^2\beta^2$, we see that $c_1=c_3^2=c_3c_4=c_4^2$.  On the other hand,
since $c_2(\alpha^2\beta+\alpha\beta^2)=c_3^2c_4\alpha^2\beta+c_3c_4^2\alpha\beta^2$, we see that $c_2=c_3^2c_4=c_3c_4^2$.  It follows that
$c_2^2=c_3^3c_4^3=c_1^3$ which does not hold for general $c_1$, $c_2$.

The conclusion is that $\alpha$ and $\beta$ must in fact
be linearly dependent
in the complex vector space 
$\Gamma(\mathcal{O}_D(-2K_D+N_{D/B}))$, so both of them can be written
as constant multiples of a section $\hat{u}$.  It follows that $\hat{f}=A\hat{u}^2$ and $\hat{g}=B\hat{u}^3$ for some constants $A$ and $B$.

Suppose that $\dim \Gamma(\mathcal{O}_D(-2K_D+N_{D/B})) > 1$.
Then not every element of $\Gamma(\mathcal{O}_D(-4K_D+2N_{D/B}))$
is the square of an element of $\Gamma(\mathcal{O}_D(-2K_D+N_{D/B}))$,
so the generic $\hat{f}$ is not of the form $A\hat{u}^2$ for any 
$\hat{u}$.  Thus. if the gauge algebra is $\mathfrak{so}(8)$, 
$\dim \Gamma(\mathcal{O}_D(-2K_D+N_{D/B}))$ must be $1$.

Suppose that $\dim \Gamma(\mathcal{O}_D(-2K_D+N_{D/B}))=1$ with
generator $\hat{u}$,
but $\dim \Gamma(\mathcal{O}_D(-4K_D+2N_{D/B})) > 1$ 
(respectively $\dim \Gamma(\mathcal{O}_D(-6K_D+3N_{D/B})) > 1$).
Then the generic element of $\Gamma(\mathcal{O}_D(-4K_D+2N_{D/B}))$
does not have the form $A\hat{u}^2$ (respectively, the generic
element of $\Gamma(\mathcal{O}_D(-6K_D+3N_{D/B}))$ does not have the
form $B\hat{u}^3$), so the maximally Higgsed
gauge algebra is not $\mathfrak{so}(8)$.

It follows that any maximally Higgsed model with $\mathfrak{so}(8)$ gauge
symmetry must take the form of Solution 1 above.

Consider now the case that the gauge algebra is $\mathfrak{so}(7)$, and
let $\lambda\in \Gamma(\mathcal{O}_D(-2K_D+N_{D/B}))$, $\mu\in \Gamma(\mathcal{O}_D(-4K_D+2N_{D/X}))$ be sections such that $\hat{f}=\mu-\lambda^2$
and $\hat{g}=-\lambda\mu$.  We cannot have
$\mu\equiv0$ or else $\hat{f}$ would be a perfect square and
the gauge algebra would be $\mathfrak{so}(8)$.
  So $\mu\not\equiv0$.
If in addition $\lambda\not\equiv0$,
then the locus $\{\hat{g}=0\}$ is the union of $\{\lambda=0\}$ and
$\{\mu=0\}$.  Thus, if we scale the coefficients $(\hat{f},\hat{g})\to
(c_1\hat{f},c_2\hat{g})$, the new sections $\lambda'$ and $\mu'$ must
satisfy $\lambda'=c_3\lambda$, $\mu'=c_4\mu$.  It follows that
$c_2=c_3c_4$ and $(c_1-c_4)\mu-(c_1-c_3^2)\lambda^2=0$.  If $\mu$
and $\lambda^2$
are linearly independent, then $c_1=c_4=c_3^2$ which
implies that $c_2^2=c_3^2c_4^2=c_1^3$, but this is
not  true for general $c_1$, $c_2$.  Thus, $\mu$ and $\lambda^2$  must be linearly dependent
and so $\lambda^2 = K\mu$ (since $\mu\not\equiv0$).  But if $K\ne0$,
then $\mu-\lambda^2=(\frac1K-1)\lambda^2$ is a perfect square, which
would force the gauge algebra to be $\mathfrak{so}(8)$.  Thus,
$K$ must be $0$ and $\lambda$ must vanish identically, and we
have solution 2 as above.
Thus, we find that a non-Higgsable $\gso_7$ is only possible when
$\hat{g}_3 = 0$ identically, and $\hat{f}_2$ is not a perfect square.
This is only possible for generic choices of $\hat{f}_2$
when either $\hat{f}_2$ contains only a single non-even monomial in a
local coordinate system, or contains multiple independent monomials.
Note that this implies that for a maximally Higgsed $\gso_7$ algebra,
$g$ must vanish to order at least 4 (rather than order 3) along the
gauge divisor.

\end{document}